# Quantum Interference and Contact Effects in Thermoelectric Performance of Anthracene-Based Molecules

Joseph M. Hamill[a*,†], Ali Ismael[b*,†], Alaa Al-Jobory[b,c,†], Troy L. R. Bennett[d,†], Maryam Alshahrani [b,e,†], Xintai Wang[b,f], Maxwell Akers-Douglas[d], Luke A. Wilkinson[d], Benjamin J. Robinson[b], Nicholas J. Long[d], Colin Lambert [b*], and Tim Albrecht [a*]

[a]School of Chemistry, University of Birmingham, Edgbaston Campus, Birmingham B15 2TT, United Kingdom.
[b]Physics Department, Lancaster University, Lancaster, LA1 4YB, UK.
[c]Department of Physics, College of Science, University of Anbar, Anbar, Iraq.
[d]Department of Chemistry, Imperial College London, MSRH, White City, London, W12 0BZ, UK.
[e]Department of Physics, College of Science, University of Bisha, Bisha, KSA.
[f]School of Information Science and Technology, Dalian Maritime University, Dalian, China.

[†]These authors contributed equally to this work.
*To whom correspondence should be addressed. e-mail: k.ismael@lancaster.ac.uk; c.lambert@lancaster.ac.uk; jmh@chem.ku.dk; t.albrecht@bham.ac.uk

## Abstract

We report on the single-molecule electronic and thermoelectric properties of strategically chosen anthracene-based molecules with anchor groups capable of binding to noble metal substrates, such as gold and platinum. Specifically, we study the effect of different anchor groups, as well as quantum interference, on the electric conductance and the thermopower of gold/single-molecule/gold junctions and generally find good agreement between theory and experiment. All molecular junctions display transport characteristics consistent with coherent transport and a Fermi alignment approximately in the middle of the HOMO/LUMO gap. Single-molecule results are in agreement with previously reported thin-film data, further supporting the notion that molecular design considerations may be translated from the single- to the many-molecule devices. For combinations of anchor groups where one binds significantly more strongly to the electrodes than the other, the stronger anchor group appears to dominate the thermoelectric behaviour of the molecular junction. For other combinations, the choice of electrode material can determine the sign and magnitude of the thermopower. This finding has important implications for the design of thermoelectric generator devices, where both n- and p-type conductor are required for thermoelectric current generation.

## Introduction

Thermoelectric power generation has interesting prospects, because it is one of only a few methods to convert waste heat into electrical energy in a low-maintenance, robust device format. The basic setup of a thermoelectric generator is shown in fig. 1, including the p- and n-type semiconducting branches, the temperature gradient and the resulting Seebeck voltage $\Delta V_S$ as well as the load resistance.[1,2] However, one of the disadvantages of the technology is that the efficiency is relatively low, fundamentally due to the Carnot limit but also limitations imposed by the materials used.[3,4][ref thermoelectrics book] To this end, the material-specific figure of merit *ZT* may be defined as shown in eq. (1),

$$ZT = \frac{GS^2}{\mathrm{k}}, \quad (1)$$

where *G* is the electrical conductance, *S* is the thermopower, and *k* is the thermal conductance. Maximising *ZT* requires the simultaneous maximisation of *S* and *G*, while minimising *k* – fundamentally difficult for materials where *G* and *k* are linked via the Wiedemann-Franz law. Indeed, achieving *ZT* > 1 at room temperature has proven to be challenging and only in recent years materials with larger *ZT* values have been found.[5,6,7] These recent breakthroughs are typically achieved through careful nanostructuring of known materials,[8,9,10] but there is a strong need for the discovery of new ones with significantly better performance. Indeed, organic materials have been identified as promising candidates, with evidence suggesting that *G*, *S* and *k* may be optimised independently to some degree, at least in some charge transport regimes.[11,12,13,14,15,16,17,18]

An additional aspects is that many thermoelectric materials, old and new, contain elements on the EU's list of critical at-risk raw materials – e.g. bismuth, hafnium, and antimony.[19] This list was compiled as part of work by the EU to identify those elements that were at-risk of becoming scarce as a result of e.g. supply chain break down. In the cases of bismuth and antimony, this risk is in part as a consequence that up to 80% of the world's supply comes from only one country, and, as with bismuth it is rarely recycled or recyclable. Telluride is a common component of many of these materials, and although it is not on this list, it is as rare as platinum in Earth's crust.[5] Organic molecules, on the other hand, may be synthesized using sustainable feed stock, which is not threatened by supply chain stresses, and with greater design flexibility beyond transition metal crystalline geometries. However, for organic thermoelectric power generation to become viable, it has been argued that eight milestones must be met.[20,21] The first four of which are to achieve:

1) a power factor $GS^2 > 10^4$ aW $K^2$
2) a phonon thermal conductance $k < 10$ pW $K^{-1}$
3) reproducible predictions and measurements of Seebeck coefficients, electrical and thermal conductances for systems with thermoelectric figures of merit ZT > 3
4) achieve comparable single-molecule and small-area predictions and measurements

The remaining Milestones are concerned with scaling-up the achievements of the first four Milestones. Our present study mainly relates to Milestones 1, 3 and 4, based on a quick and effective method for characterizing single-molecule electric conductance and Seebeck coefficients.[22] Here, we set out to utilize this method to explore their dependence on molecular connectivity and anchor groups for a set of anthracene-based molecules, fig. 2, which are known to feature quantum interference effects.[23,24] Supported by theoretical calculations, and by comparing to previously reported thin-film results, we explore the effect of the nature of the anchor groups in combination with the substrate material. Interestingly, apart from variations in the magnitude of the Seebeck coefficient, we have observed a sign reversal resulting from a change in junction from Au/Pt to Au/Au. While such behaviour has been observed before, for benzenedithiol in Au/Au and Au/Ni junctions and been rationalised based on spin hybridisation at the Fermi level,[25] in our case, the change in sign appears to reflect a more subtle difference in the bonding interaction between the anchor groups and the substrate electrodes, with concomitant changes in Fermi level alignment.

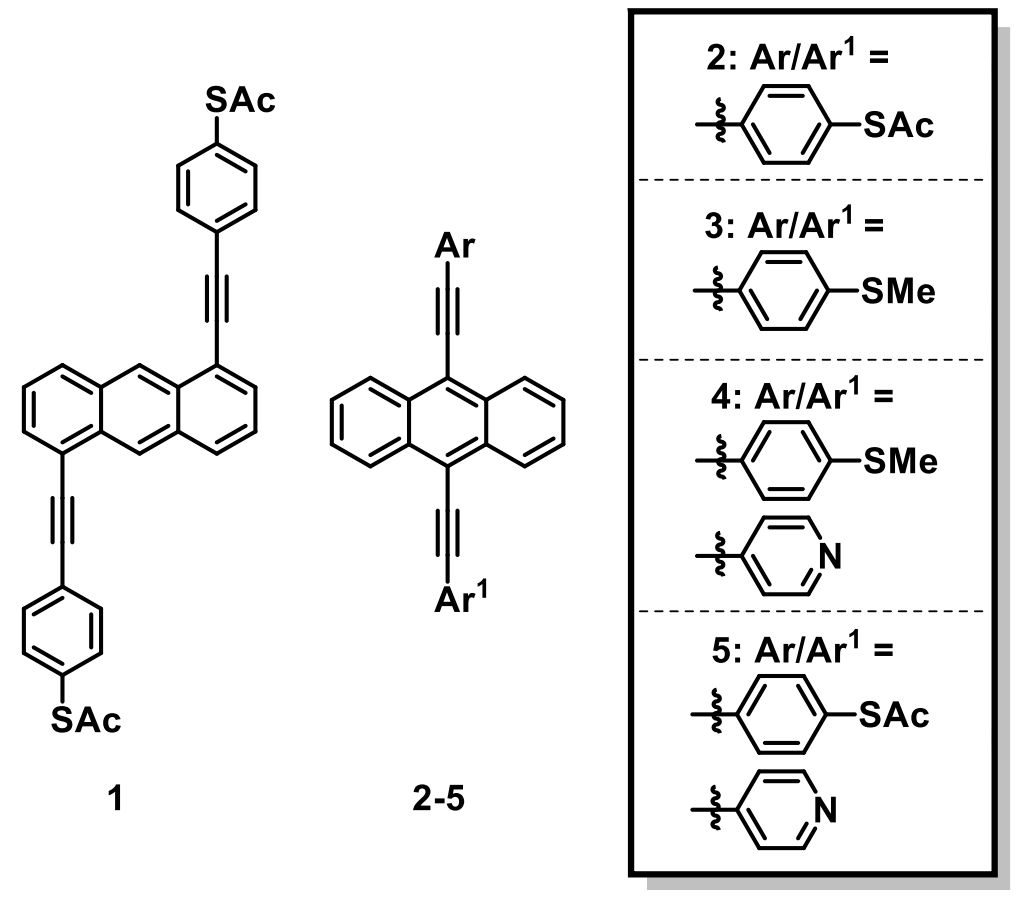

**FIG. 2:** (a) Molecules measured in this study:, 1,5-Di(4-(ethynylphenyl)thioacetate)anthracene (**1**), 9,10-Di(4-(ethynylphenyl)thioacetate)anthracene (**2**), 9,10-Di(4-ethynylthioanisole)anthracene (**3**), 9-(4-(ethynylphenyl)thioacetate)-10-(4-ethynylpyridine)anthracene (**4**), 9-(4-ethynlthioanisole)-10-(4-ethynylpyridine)anthracene (**5**).

## Experimental Section

Chemicals and Synthesis of molecules **1**-**5**, fig. 2. We have previously reported the synthesis of compounds **1**-**4** and refer the reader to reference 23 for further details of their synthesis and characterisation. Compound **5** was synthesised by employing a stepwise Sonogashira methodology utilising reactions between 9,10-dibromoanthracene and terminal alkynes. 4-(Ethynylphenyl)thioacetate can undergo a self-oligomerisation to form a cyclic trimer when exposed to Sonogashira conditions.[26] In order to avoid this unwanted side reaction we decided to utilise a protecting-group strategy. Our previous work utilised a *tert*-butyl protecting group which could be interconverted to a thioacetate through treatment with boron tribromide to allow for dealkylation, followed by quenching with acetic anhydride. In our experience, however, attempts to apply this methodology in the synthesis of compound **5** were unsuccessful. Considering this, we moved to the use of a cyanoethyl protected thiol, which presents much milder deprotection conditions. To this end, we synthesised 4-(ethynylphenyl)-thiocyanoethyl following the methodology presented by *Bryce et al*. and subsequently reacted this with 9,10-dibromoanthracene in a 1:5 ratio under Sonogashira conditions.[27] This reaction generated a mixture of the monosubstituted **(5A**, see SI S1.2) and symmetrically disubstituted products (**5B**) which could be trivially separated from one-another using flash chromatography. The monosubstituted product was subsequently reacted with 4-ethynylpyridine under analogous conditions to produce an asymmetrically disubstituted product (**5C**). The final step involved interconversion of the thiol protecting group through first treating compound (**5C**) with sodium methoxide to allow for removal of the cyanoethyl group before quenching with acetic anhydride to generate a terminal thioacetate. This was purified using an aqueous work-up to provide compound (**5**) in good yield. Further details can be found in sec. S1 of the SI.

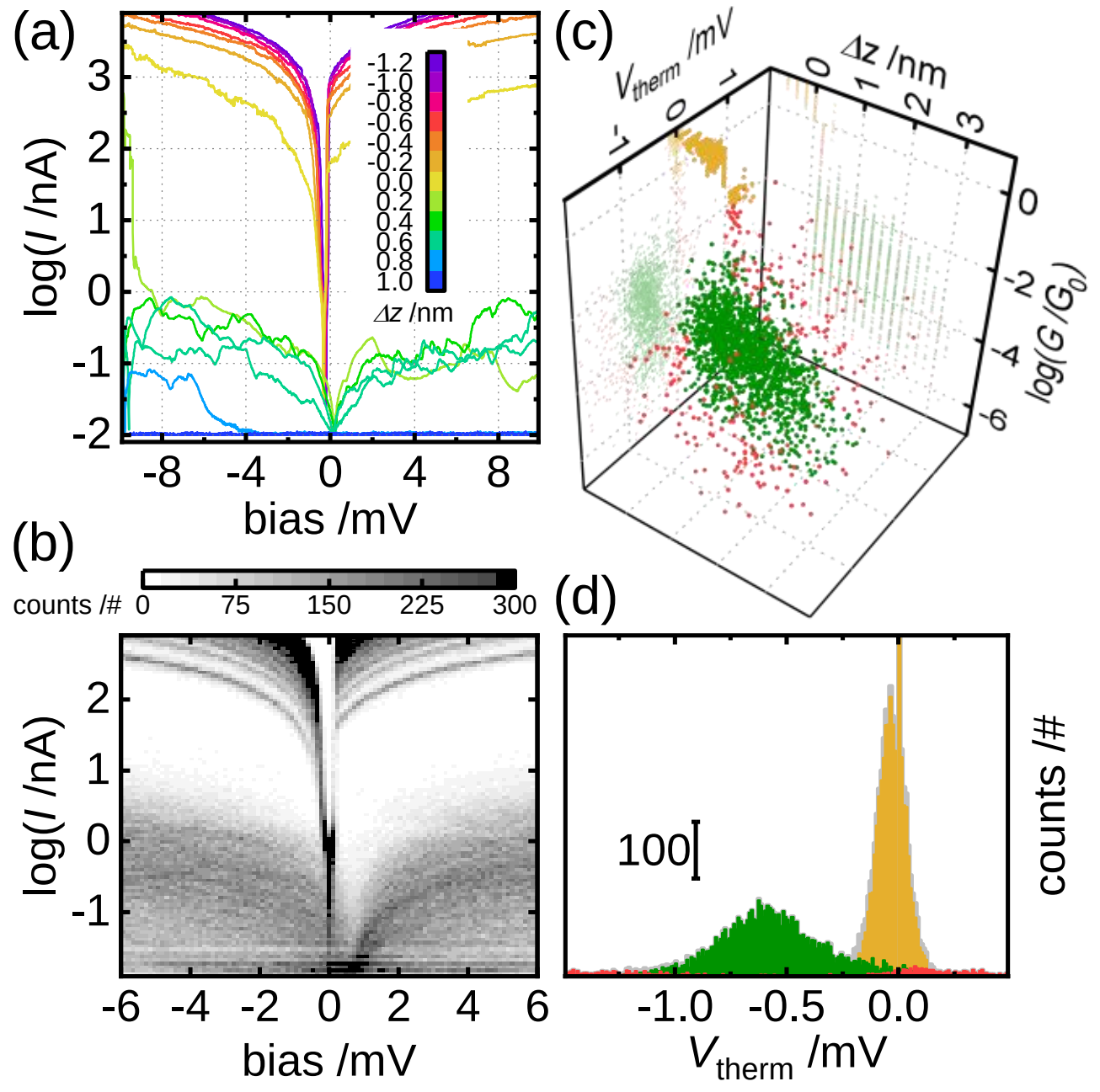

**FIG. 3**: Example STM IV experiment performed on adlayer of molecule **2**. (a) Example withdraw series with I/V sweeps across the Au/Au junction in purple through yellow – yellow sweep is the last Au/Au sweep. Sweeps across Au/**5**/Au in green. Sweeps across the open junction in blue. (b) 2D current vs bias intensity plot after $V_{corr}$ is removed from each group. (c) 3D scatter plot of displacement, $\Delta z$, conductance, $G$, and voltage offset, $V_{therm}$, from each trace (sweeps across the open junction are removed) clustered using Gaussian mixture model into a Au/Au cluster (gold) and a molecular cluster (green), and a noise cluster (red). (d) 1D histograms of $V_{therm}$ for the three clusters above, and for the entire data set (grey) at $\Delta T$ = 27 K.

For the determination of single-molecule Seebeck coefficients, a distance-dependent scanning tunnelling microscopy (STM) current-voltage (I/V) method was used.[22] Briefly, the tip was first brought into contact with the substrate surface and then withdrawn in 25 steps of 0.2 nm. During each step, the bias voltage was swept between ±10 mV at a rate of 0.2 V $s^{-1}$ and the current recorded (tip withdrawal rate: ~2 nm $s^{-1}$, 2.5 s per series), cf. Fig. 3(a) as an example. Typically, three different classes of *I*/*V* sweeps were observed, namely Au/Au (purple to yellow), Au/molecule/Au (green) and open junctions (blue). The conductance *G* for each *I*/*V* sweep was determined, based on 41 data points centred at -5 mV. The sweep with a conductance closest to, but larger than the quantum conductance $G_0$ was taken to define the voltage correction for each *I*/*V* sweep within a given series. At each $\Delta T$, *ca*. 1000 withdrawals and thus 25000 *I*/*V* traces were recorded. The sweeps were parameterized into the three-dimensional space ($\Delta z$, $G$, $\Delta V$), Fig. 3(c), and clustered into three clusters using a Gaussian Mixture Model.[28,29] To illustrate the voltage shift due to $\Delta T$, the 1D histogram of $\Delta V$ values for molecule **2** at $\Delta T$ = 27 K is shown in Fig. 3(d): sweeps assigned to Au/Au junctions (yellow) are tightly centred around 0 µV, sweeps assigned to noise (red) widely distributed, while sweeps assigned to Au/molecule/Au sweeps (green), show a clear offset of 0.5 mV. For each molecule, experiments were conducted at 4-10 different $\Delta T$ values and results for each analyte replicated on different days. Each of the three parameters were plotted vs $\Delta T$, Fig. 4, and each replicate fitted separately (light and dark red lines in Figs. 4(a)-(c)). A combined linear fit was calculated to determine a overall slope and standard error of the slope, and plotted along with 95% confidence interval, fig. 4 (d), for the voltage correction (at Au/Au contact)[22] as well as molecules **1**, **2**, **4** and **5**. Fig. 4 (e) shows $S_{mol}$ for each replicate (blue/red/green), and an overall $S_{mol}$ (orange) (error bars: standard error of slope), cf. also fig. S9, S10 in the SI.

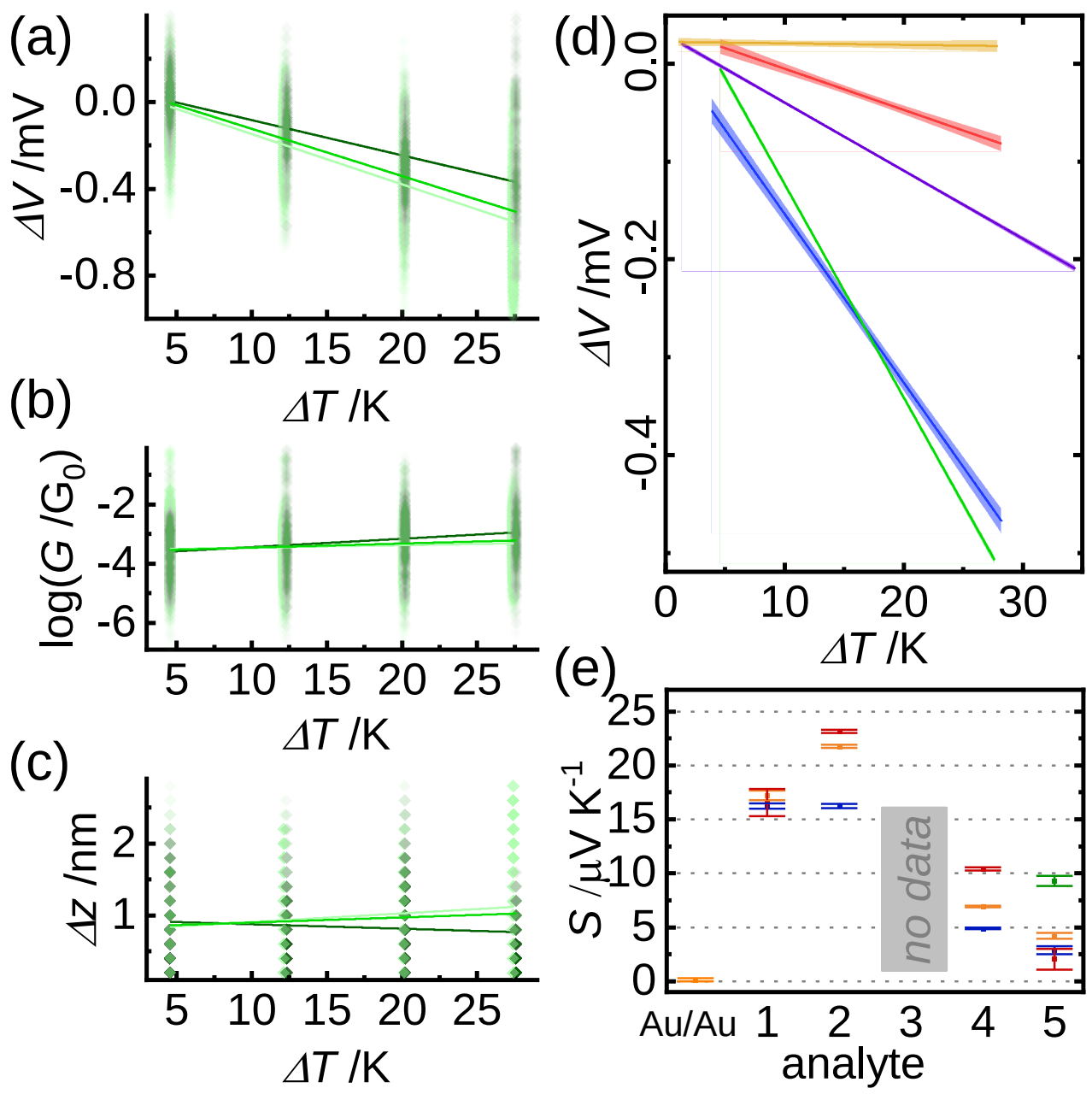

**FIG. 4.** (a) Scatter plots of $\Delta V$ vs $\Delta T$ measurements of molecule **2** with separate trend lines for two separate experiments (light and dark green), and combined trend lines with 95% confidence intervals (green). (b) Scatter plots of $G$ vs $\Delta T$, and (c) $\Delta z$ vs $\Delta T$ for the same measurements, with $G_{mol}$ and $\Delta z_{mol}$ from each separate measurement calculated as the mean of a Gaussian fit of all data, and standard deviation as error bars. Trend lines are aids for the eye. (d) $\Delta V$ vs $\Delta T$ trend lines with 95% confidence intervals for molecules **1** (blue), **2** (green), **4** (purple), **5** (red) and clean Au/Au (gold). (e) Summary of $S_{mol}$ for all molecules in this study, and the internal reference at Au/Au contact. Blue/red/green represent trial 1/2/3, and orange is the combined result (error bars: standard error of the slope from the linear least-square fit).

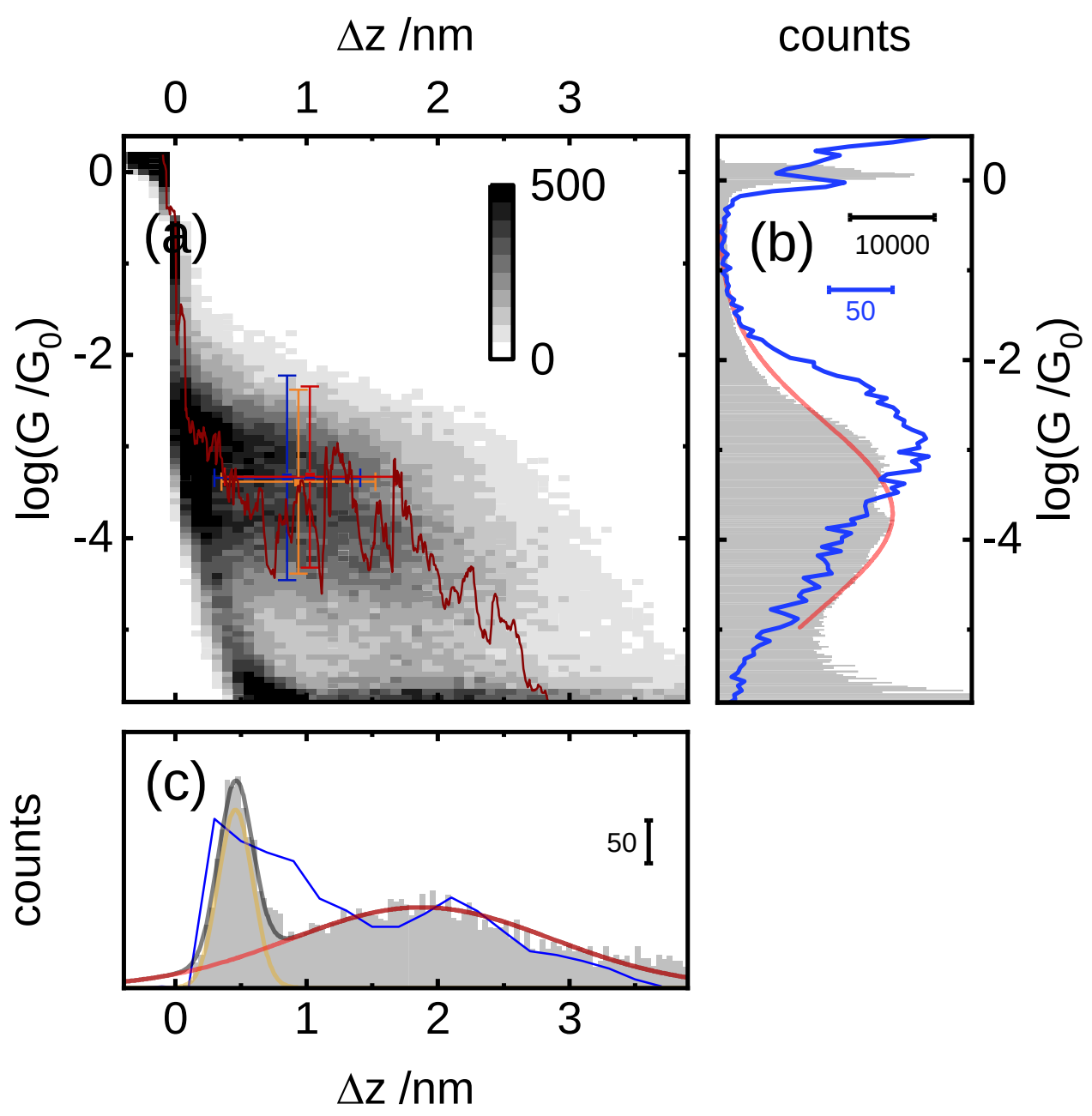

**FIG. 5**: Example constant bias STM BJ experiment performed on adlayer of molecule **2**. (a) *ca.* 7k traces combined in a 2D conductance vs displacement intensity plot with an example trace (crimson). Crosshairs are ($\Delta z_{mol}$, $G_{mol}$) from the STM IV measurements with standard deviation (trials 1 and 2 (blue), combined dataset (orange)). (b) 1D conductance histogram of all traces from constant bias measurements in grey and Gaussian fit of molecular plateau region (red). Blue: 1D conductance histogram of all sweeps from STM IV mode from data plotted in Fig. 3(c). (c) 1D displacement histogram of all traces in constant bias measurement, determined at conductance of $10^{-5.2}$ $G_0$. Two-peak area-type Gaussian fit (red/yellow). Junction formation probability: 77%. 1D displacement histogram from molecular cluster in STM IV mode (blue line), from data plotted in Fig. 3(c).

Constant bias STM BJ measurements were also performed on all analytes, to compare with results from STM IV measurements and to potentially gain additional insight into the junction geometry and progression.[22] In brief, the STM tip is initially brought into contact with the substrate surface at a constant tip/substrate bias (here: 100 mV). It is then withdrawn at a constant rate, typically between 8-16 nm $s^{-1}$, and the current recorded. A typical withdraw trace is plotted in Fig. 5(a) in crimson for a measurement of molecule **2**, showing the range from Au/Au contact to noise level. The region where charge transport through the molecular bridge dominates is indicated by a plateau-like region. The intensity plot of *ca.* 7400 traces in Fig. 5(a) exhibits both tunnelling traces (“empty” gaps) and molecular traces. The tunnelling traces are evident by the dense cloud of short traces which decay linearly between $10^{-4}$ and $10^{-5}$ $G_0$. The molecular traces show more variation and exhibit a broadly distributed plateau region at $10^{-3.5}$ $G_0$ that extends for about 2 nm. The mean conductance $G_{mol}$ of the molecular plateau was determined from a Gaussian fit (red) of 1D histogram of conductance values, Fig. 5(b). The tunnelling traces contributed negligibly to the conductance histogram, because they exhibit few data points in the molecular region. To determine the plateau length, distance between G = $G_0$ and G ≤ $10^{-5.2}$ $G_0$, was determined for each trace. A histogram of the plateau lengths, in grey in Fig. 5(c), yielded two peaks. The first peak at *ca.* 0.5 nm was due to rapidly decaying tunnelling traces (yellow), while the second peak (red), at *ca*. 2.0 nm, is related to molecular junctions (77%, based on the relative area of the Gaussian fits. For comparison, $\Delta z_{\mathrm{mol}}$ and $G_{\mathrm{mol}}$ values from STM IV measurements (replicates in red/blue, combined dataset in orange; error bars: standard errors of the means). Hence, the close mapping of the STM IV results onto the STM BJ results was a strong confirmation that the molecules were present in the STM IV measurements, and that the clustering step of the analysis was selecting for molecular I/V sweeps. See Figs. S11-S13 in the SI for STM BJ results for molecules **1**, **4**, and **5**).

The transport properties of the studied junctions were further investigated using a combination of density functional theory and quantum transport theory[30] to obtain the transmission coefficient $T(E)$ describing electrons of energy $E$ passing from the source to the drain electrodes[31]. Using the density functional code SIESTA the optimum geometries of isolated molecules were obtained by relaxing the molecules until all forces on the atoms were less than 0.01 eV / Å.[32,33] A double-zeta plus polarization orbital basis set, norm-conserving pseudopotentials, an energy cut-off of 250 Rydbergs defining the real space grid were used and the local density approximation (LDA) was chosen as the exchange correlation functional. We also computed results using GGA and found that the resulting transmission functions were comparable with those obtained using LDA.[34,35,36] To calculate the optimum binding distance between a molecule and electrode, we used DFT and the counterpoise method, which removes basis set superposition errors (BSSE). The binding distance *d* is defined as the distance between the molecule A and electrode B. The ground state energy of the total system is calculated using SIESTA and is denoted $\mathrm{E}_{\mathrm{AB}}^{\mathrm{AB}}$. The energy of each entity is then calculated in a fixed basis, which is achieved using ghost atoms in SIESTA. Hence, the energy of A in the presence of the fixed basis is defined as $\mathrm{E}_{\mathrm{A}}^{\mathrm{AB}}$ and for the electrode B as $\mathrm{E}_{\mathrm{B}}^{\mathrm{AB}}$. The binding energy is then calculated using the following equation: $BE = E_{AB}^{AB} - E_{A}^{AB} - E_{B}^{AB}$. Transmission coefficient curves *T(E)* were obtained using the GOLLUM transport code.[30] Following this, the Seebeck coefficient (S) of the junction was calculated as described in section 3 of the supplementary information.

## Results and Discussion

The results for molecules **1**-**5** are summarized in Table 1, for both STM IV and STM BJ methods (top and bottom values in each row), and column-wise from left to right, $\Delta z_{mol}$, $G_{mol}$, $S_{mol}$ and the power factor *f*. We note that for **3**, we were unable to obtain reproducible results using the STM IV method and hence no thermopower value could be determined. The final row represents nominal results from "empty" tunnelling junctions, i.e. in the absence of a molecular bridge, as described in further detail in ref 22. The data lend themselves to several broad observations: 1) $\Delta z_{mol}$ values are usually found to be close to 1 nm or just below, which is shorter than the value of approximately 2 nm expected for fully extended bridges of these molecules. Exceptions are the values determined for **1** and **2** using STM BJ, where the $\Delta z_{mol}$ values are in good agreement with theoretical expectations. Both molecules feature thiol-based anchor groups, which form strong bonds to the respective gold electrode contacts. This suggests that for **3**-**5**, the weaker interaction between the SMe- and N anchors with the electrode substrates may limit the lifetime of the junction. Weaker than the gold-thiol bond, recent XPS studies indicated that their interaction with the gold substrate may be of comparable strength, based on the distribution of anchor groups on the Au surface, [37] even our theoretical findings do show some difference, *vide infra*. 2) With regards to $G_{mol}$, both spectroscopic methods yielded the same values for **4** and **5**, within experimental error. For **1** and **2**, however, STM BJ yielded smaller $G_{mol}$ values, by an order of magnitude, compared STM IV, and in conjunction with the longer break-off distance, this might suggest a non-negligible contribution from other conductance pathways, including "through-space" tunnelling.[38] Comparing $G_{mol}$ for **1** and **2** for the same spectroscopic method, we find the value for **2** to be about one order of magnitude larger than for **1**, broadly in line with expectations from Magic Number theory and highlighting the effect of quantum interference on the junction conductance. None of the molecules appears to be particularly conductive though, with $G_{mol}$ values smaller than -3, in logarithmic units of $G_0$. 3) The $S_{mol}$ values were determined successfully for **1**, **2**, **4** and **5**, where those for **1** and **2** are similar and significantly higher than for **4** and **5**. This could suggest that in the latter two cases, the Fermi level is closer to the centre of the HOMO/LUMO gap. The magnitudes $G_{mol}$ and $S_{mol}$, and hence the power factor $f$, are however small for all molecules studied here, in comparison to the Milestones listed above. Even for the best performing molecule **2**, the value of $f = 16$ is still significantly below the stipulated value of $10^4$, thus requiring further optimisation of both $G_{mol}$ and $S_{mol}$, for example by careful design of the electronic structure of the junction.

Finally, all molecules showed positive $S_{mol}$ values, suggesting that charge transport is HOMO-dominated and likely due to the sulphur-based anchor groups. Bar one exception, the magnitude of $S_{mol}$ is comparable to previously reported values for molecules **1**, **2** and **5**, determined in Au/Pt thin-film junctions, see references 23,24 and fig. 6. The exception is **4**, where we find $S_{mol} > 0$, while previous work in Au/Pt thin-film devices yielded $S_{mol} < 0$. This would imply a change in charge transport mechanism from hole-dominated to electron-dominated transport and may be induced by a slight shift of the Fermi level offset, e.g. due to differences in the interaction between the anchor groups and the respective substrate materials (Au/Au vs. Au/Pt).

**Table 1.** Results from current-distance spectroscopy at constant bias (top) and distance dependent *I/V* spectroscopy (bottom). Values for $\Delta z_{\mathrm{mol}}$ and $G_{\mathrm{mol}}$ are sample mean and standard deviation for all $\Delta T$ and all replicates, *S* values are slope and standard error of the slope for the trend line through all $\Delta T$ and all replicates.

| *Molecule* | $\Delta z_{\mathrm{mol}}$ nm | $G_{\mathrm{mol}}$ log(G $G_0^{-1}$) | $S_{\mathrm{mol}}$ µV $K^{-1}$ | *f* aW $K^{-2}$ |
|---|---|---|---|---|
| **1** (1,5 SAc2) | 2(1) | -5(1) | | |
| | 0.8(5) | -4(1) | 17.2(4) | 4.5 |
| **2** (9,10 SAc2) | 2(2) | -4(2) | | |
| | 0.9(6) | -3(1) | 21.8(1) | 16 |
| ***3*** *(*9,10 SMe2*)* | *1(1)* | *-3(2)* | | |
| **4** (9,10 SMe,N) | 1.1(3) | -4.0(9) | | |
| | 0.8(4) | -4(1) | 6.95(7) | 0.70 |
| **5** (9,10 SAc,N) | 1(1) | -4(1) | | |
| | 0.7(5) | -4(1) | 4.2(3) | 0.41 |
| ***Au/Au*** | 0.4(2) | -3(1) | (-0.2(2)) | |

To explore the electronic structure of the junctions and the effects of substrate metal and anchor groups on $S_{mol}$ in more detail, we undertook a detailed DFT study, see section 3 in the SI for details, which led to the following main conclusions. Firstly, the simulations show that thiol-terminated anthracene binds about two times stronger to an Au electrode than a pyridyl/SMe-terminated anthracene (with binding energies approximately 1.0 eV vs. 0.5 eV), cf. fig. S16 and figs. S17-S22 for the optimised structures of the respective junctions. Secondly, as expected for anthracene-based polyaromatics, we found that closest agreement between theory and experiment is obtained when the Fermi energy is near the middle of the HOMO/LUMO gap, fig. S22-S26. The corresponding Seebeck coefficient as a function of Fermi level offset for molecules **1**-**5** is shown in figures S27-S31, for comparison. This suggests that off-resonance charge transport is dominant, in line with the observed electric conductance values, and also that relatively subtle changes in the electronic structure of the junction could move the Fermi level in a way that leads to a switch from HOMO- to LUMO-dominated transport or vice versa. This seems to be the case for molecule **4**, where we obtained a small, but positive $S_{mol}$, while the latter was found to be negative in previous thin-film studies in Au/Pt junctions.[23,24] Hence, simulations investigating the difference(s) between those two electrode configurations thus led us to the third conclusion, as outlined below. To this end, we simulated the two experimental setups for molecules **1**, **2** and **4**, i.e. where both experimental datasets are available. Transmission curves for **1** and **2** with Au-Au and Au-Pt electrodes are similar, even though one difference appears to be that for Au-Pt junctions, the frontier orbitals are downshifted towards lower energies by about 0.2 eV, as shown in figs. S32 and S34, likely reflecting the different electron affinities of the two metals (Au = 223 kJ/mol, Pt = 205 kJ/mol), see section 3.7 in the SI. Accordingly, molecules **1** and **2** feature positive $S_{mol}$ values in both electrode configurations, as shown in figs. S33 and S35.

However, for molecule **4**, the S-Au interaction is via a weaker SMe anchor, which does not dictate the electronic structure of the junction in the same way. As a result, the change from Au/Pt to Au/Au substrate electrodes leads to a downward shift of the transmission function, relative to $E_F$, thereby switching charge transport from HOMO-dominated ($S_{mol} < 0$) to LUMO-dominated ($S_{mol} > 0$). Crucially, it appears that the absence of a dominating anchor group allowed for this subtle effect to be observable.

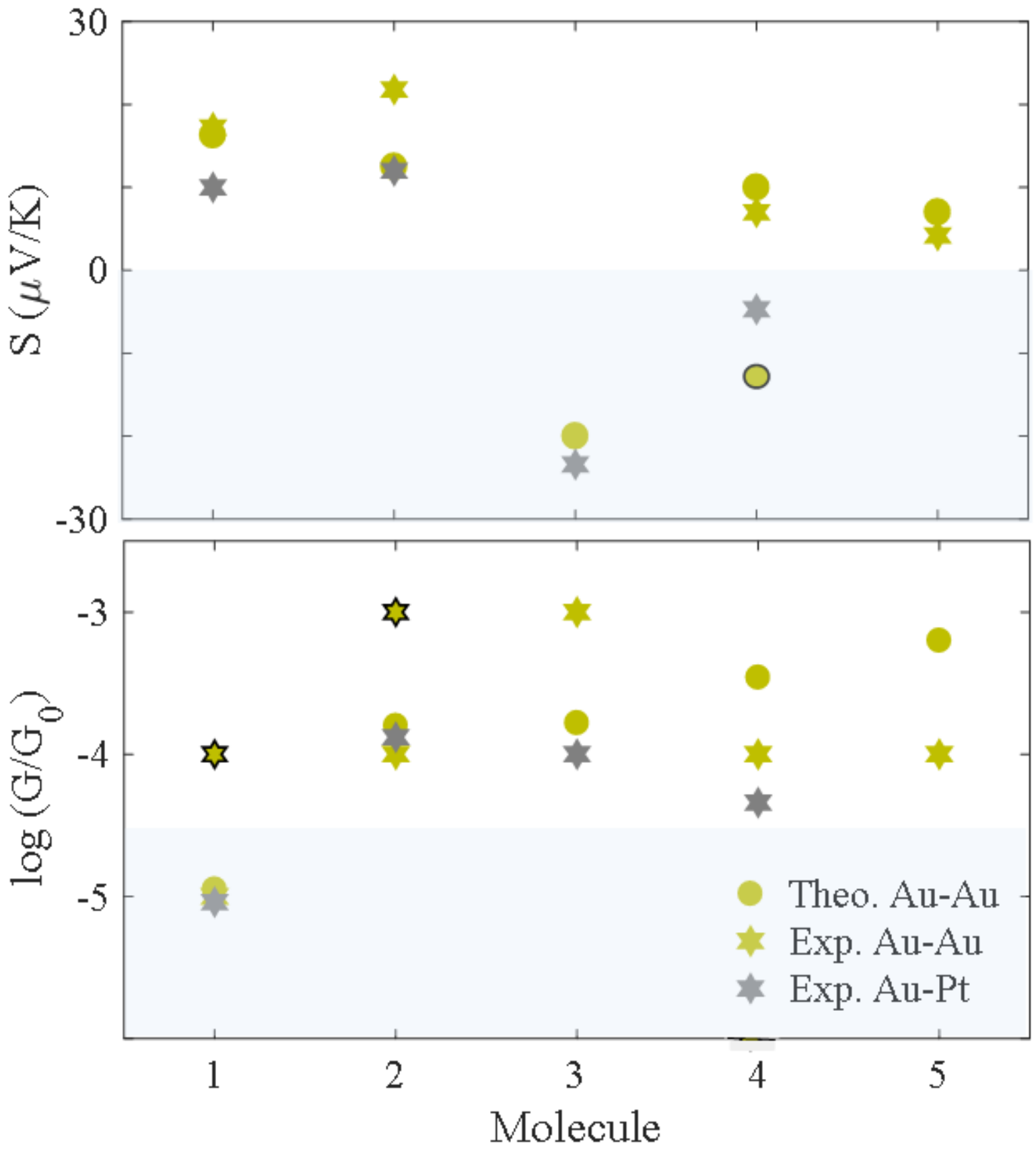

**FIG. 6**: Electric and thermoelectric properties of **1**-**5**. A comparison between experiment and theory. Bottom panel: $G_{mol}$ values from theory in Au-Au and Au-Pt junctions for **1-5** are similar at DFT-predicted mid-gap (green circles). Experiments for **1** and **2** (present study) formally yield a high $G_{mol}$ value (black-green stars) and a more representative low one (green stars), see above. Experimental $G_{mol}$ values in Au-Pt junctions for **1-4** (grey stars).[23],[24] Top panel: $S_{theory}$ in Au-Au junctions for **1, 2, 4** and **5** are positive, whereas for **3** it is negative (green circles). Similarly, $S_{theory}$ in Au-Pt junctions for **4**, is negative (black-green circle). *S* experiment in Au-Au junctions for **1, 2, 4** and **5**, are all positive (green stars). $S_{exp}$ in Au-Pt junctions for **1** and **2** are positive while **3** and **4** are negative (grey stars). Note: Experimental $S_{mol}$ values are unavailable for **3** and **5** in Au-Au and Au-Pt junctions respectively, theoretical values are all mid-gap simulations ($E_F$-$E_F^{DFT}$ ≈mid-gap).

## Conclusions

In conclusion, the present study has revealed a range of new insights into the electric and thermoelectric properties of molecular junctions, where charge transport appears to occur in the off-resonant coherent tunnelling regime. We provide a detailed comparison of two methodologies for the measurement of single-molecule charge transport, the well-established STM BJ technique at constant tip/substrate bias and distance-dependent STM IV spectroscopy, STM IV. To this end, detailed analysis revealed how, under the experimental conditions used, both methods yielded shorter than expected break-off distances, based on the length of the fully extended molecular junction. The exception were molecules **1** and **2** in STM BJ experiments, where the measured $\Delta z_{mol}$ values correspond well with theoretical expectations. While some of the apparent decrease of $\Delta z_{mol}$ in STM IV spectroscopy may be due to the limited spatial resolution of the measurement (step size: 0.2-0.3 nm), this is not sufficient to explain the observed differences, which are on the order of 1 nm or so. Notably, the applied tip/substrate bias is smaller in STM IV experiments than in STM BJ measurements (± 10 mV vs. 100 mV), so current- or heating-induced effects are also unlikely to constitute a satisfactory explanation. Since the recording of a withdrawal series in STM IV takes somewhat longer than a withdrawal in STM BJ, it is possible that mechanical drift effects lead to an on average earlier junction rupture, a hypothesis that would require further systematic study, but is beyond the scope of the present work.

Further significant improvements in $G_{mol}$ and $S_{mol}$ are however required to reach more meaningful performance characteristics, which is a reflection of the non-resonant, “mid gap” nature of charge transport through the junction. However, our results further support the notion that quantum interference effects can be harnessed to increase $G_{mol}$, as observed for molecules **1** and **2**, and potentially also $S_{mol}$. Interestingly, we find that for the molecular systems studied here, a strong imbalance between the anchor groups and their interaction with the electrode substrate, can lead to a “pinning” effect, where the stronger anchor group effectively dictates the Fermi alignment and hence the nature of the dominating charge carriers. Where such imbalance is not present, for example in molecule **4**, subtle differences in bond strength between the anchor and different substrates can lead to a change in Fermi level alignment and a switch from electron to hole transport or vice versa. Overall, $S_{mol}$ values determined from single-molecule measurements appear to compare well with those extracted from thin-film experiments, where available. This reinforces the important role of single-molecule experiments in identifying structure-function relationships and the optimisation of molecular and interfacial structure.

## ACKNOWLEDGMENT

A.K.I. and L.A.W. acknowledge support from the Leverhulme Trust for Early Career Fellowship ECF-2020-638. The project team acknowledge financial support from the UK EPSRC, through grant nos. EP/M014452/1, EP/P027156/1 (A.K.I and C.J.L.) and EP/N03337X/1 and EP/N03337X/2. This work was additionally supported by the European Commission is provided by the FET Open project 767187 – QuIET and the EU project Bac-to-Fuel. M. A. is grateful for financial assistance from Bisha-University (Saudi Arabia), and the Saudi Ministry of Education. A.K.I and A.A are grateful for financial assistance from Tikrit and Anbar Universities (Iraq), and the Iraqi Ministry of Higher Education (SL-20).

# Quantum Interference and Contact Effects in Thermoelectric Performance of Anthracene-Based Molecules

Joseph M. Hamill[a*,†], Ali Ismael[b*,†], Alaa Al-Jobory[b,c,†], Troy L. R. Bennett[d,†], Maryam Alshahrani[b,e,†], Xintai Wang[b,f], Maxwell Akers-Douglas[d], Luke A. Wilkinson[d], Benjamin J. Robinson[b], Nicholas J. Long[d], Colin Lambert[b*], and Tim Albrecht[a*]

[a]School of Chemistry, University of Birmingham, Edgbaston Campus, Birmingham B15 2TT, United Kingdom.
[b]Physics Department, Lancaster University, Lancaster, LA1 4YB, UK.
[c]Department of Physics, College of Science, University of Anbar, Anbar, Iraq.
[d]Department of Chemistry, Imperial College London, MSRH, White City, London, W12 0BZ, UK.
[e]Department of Physics, College of Science, University of Bisha, Bisha, KSA.
[f]School of Information Science and Technology, Dalian Maritime University, Dalian, China.

†These authors contributed equally to this work.

*To whom correspondence should be addressed. e-mail: k.ismael@lancaster.ac.uk; c.lambert@lancaster.ac.uk; jmh@chem.ku.dk; t.albrecht@bham.ac.uk

## 1. Molecular Synthesis

### 1.1 Materials and Methods

All reactions were performed with the use of standard air-sensitive chemistry and Schlenk line techniques, under an atmosphere of nitrogen. No special precautions were taken to exclude air during any work-ups. All commercially available reagents were used as received from suppliers, without further purification. 4-(Ethynyl)phenyl-thiopropionitrile was synthesised through an adapted literature procedure.[1] Solvents used in reactions were collected from solvent towers sparged with nitrogen and dried with 3 Å molecular sieves, apart from DIPA, which was distilled onto activated 3 Å molecular sieves under nitrogen. We have previously reported the synthesis of compounds (**1**-**5**).[2,3] Sonogashira couplings reported within this work were performed following a previously established method.[3,4]

Instrumentation. $^{1}H$ and $^{13}C\{^{1}H\}$ NMR spectra were recorded on a Bruker Avance 400 MHz spectrometer and referenced to the residual solvent peaks of $CDCl_3$ at 7.26 and 77.16 ppm, respectively. Coupling constants are measured in Hz. Mass spectrometry analyses were conducted by Dr. Lisa Haigh of the Mass Spectrometry Service, Imperial College London.

## 1.2 Synthesis

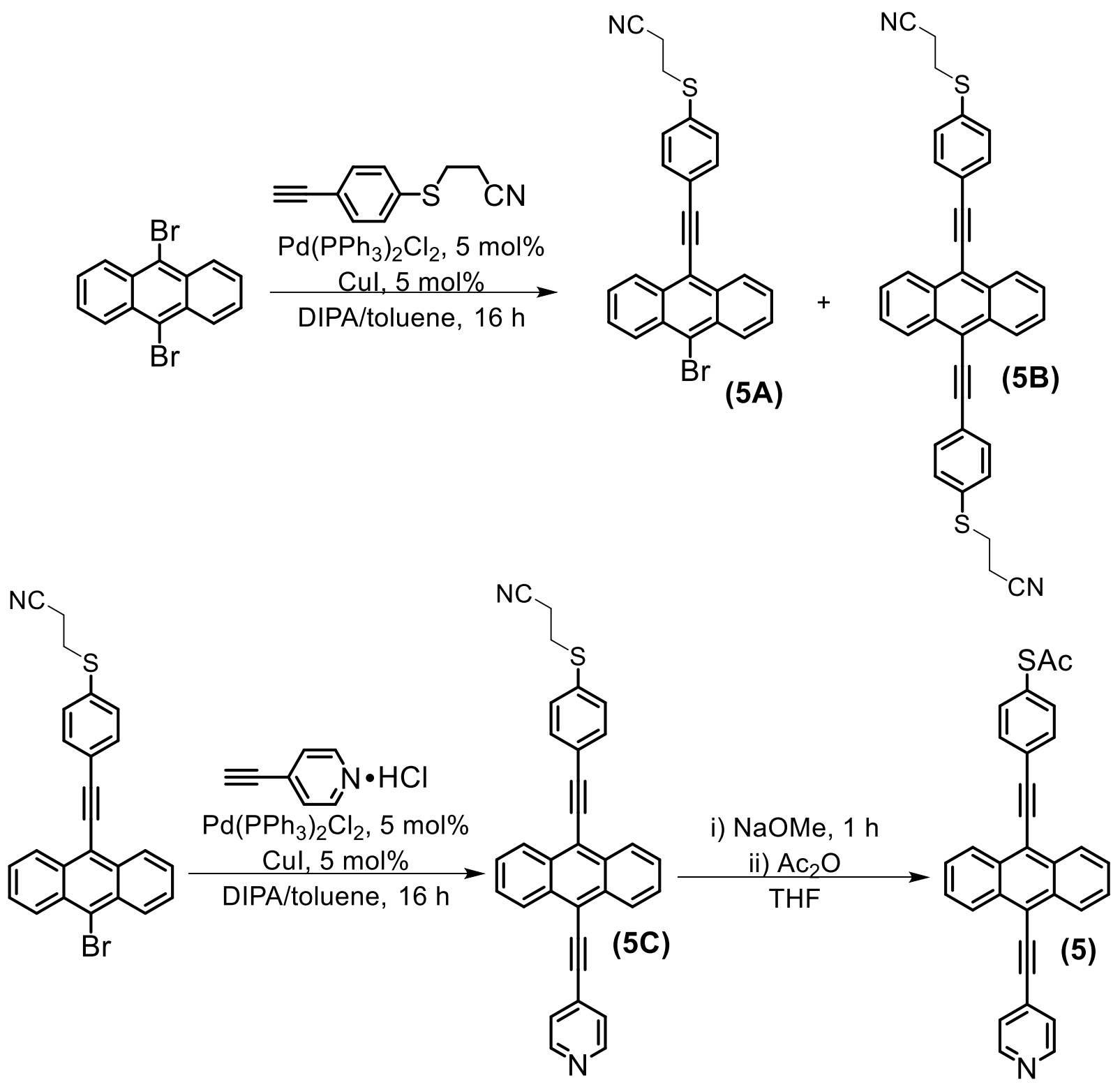

Scheme S1: Reaction pathway utilised to synthesise molecule **5**.

9-Bromo-10-(4-(ethynyl)phenylthiocyanoethyl)anthracene (8A)

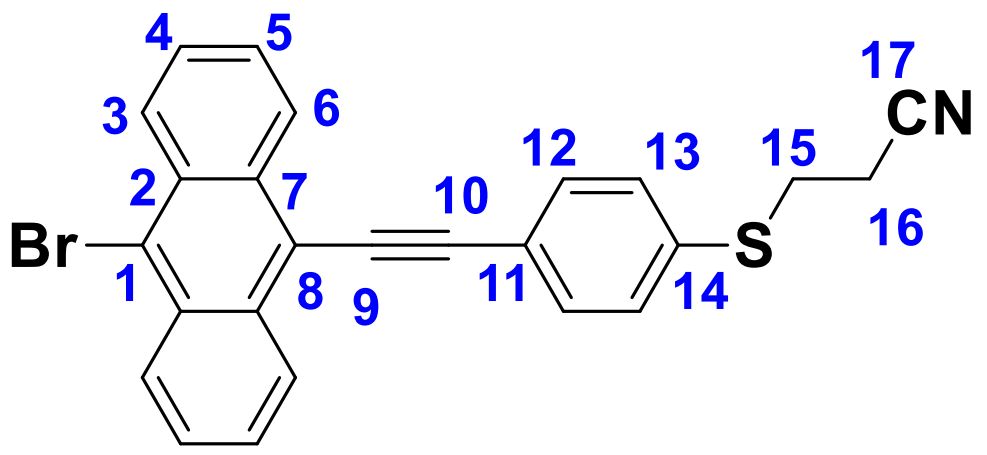

9,10-Dibromoanthracene (3.60 g, 10.72 mmol), 4-(ethynyl)phenylthiocyanoethyl (0.51 g, 2.71 mmol) and CuI (0.06 g, 0.29 mmol) were dissolved in DIPA (10 mL) and toluene (250 mL). $Pd(PPh_3)_2Cl_2$ (0.02 g, 0.33 mmol) was added and the mixture was stirred at 50°C for >16 hours to give a dark green solution. The solvent was removed *in vacuo* before the crude product was dissolved in DCM, filtered and subsequently purified by chromatography on an alumina V column, eluting with n-hexane/DCM (1:0 → 4:1) to give the product as a bright orange solid (0.52 g, 1.18 mmol, 43%).

**$^1$H NMR** ($CDCl_3$, 298 K, 400 MHz): **$\delta_H$** = 8.65-8.60 (m, 2H, *H*6), 8.58-8.53 (m, 2H, *H*3), 7.70 (dt, $^3J_{H\text{-}H}$ = 8.8, $^4J_{H\text{-}H}$ = 2.0 Hz, 2H, *H*13), 7.66-7.59 (m, 4H, *H*4, *H*5), 7.44 (dt, $^3J_{H\text{-}H}$ = 8.8, $^4J_{H\text{-}H}$ = 2.0 Hz, 2H, *H*12), 3.21 (t, $^3J_{H\text{-}H}$ = 7.6 Hz, 2H, *H*15), 2.67 (t, $^3J_{H\text{-}H}$ = 7.6 Hz, 2H, *H*16) ppm; **$^{13}$C{$^1$H} NMR** ($CDCl_3$, 298 K, 100 MHz): **$\delta_C$** = 134.6 (Ar-$\underline{C}$-C), 133.1 (Ar-$\underline{C}$-C), 132.5 (Ar-$\underline{C}$-H), 130.6 (Ar-$\underline{C}$-H), 130.4 (Ar-$\underline{C}$-C), 128.4 (Ar-$\underline{C}$-H), 127.6 (Ar-$\underline{C}$-H), 127.2 (Ar-$\underline{C}$-H), 127.1 (Ar-$\underline{C}$-H), 124.7 (Ar-$\underline{C}$-C), 122.6 (Ar-$\underline{C}$-C), 118.0 (Ar-$\underline{C}$-C), 117.9 (Ar-$\underline{C}$-C), 101.0 (-$\underline{C}$≡C-), 87.4 (-$\underline{C}$≡C-), 29.9 (-$\underline{C}H_2$-), 18.4 (-$\underline{C}H_2$-) ppm; **MS APCI**: calcd. [M]+ 442.0260; found. 442.0272.

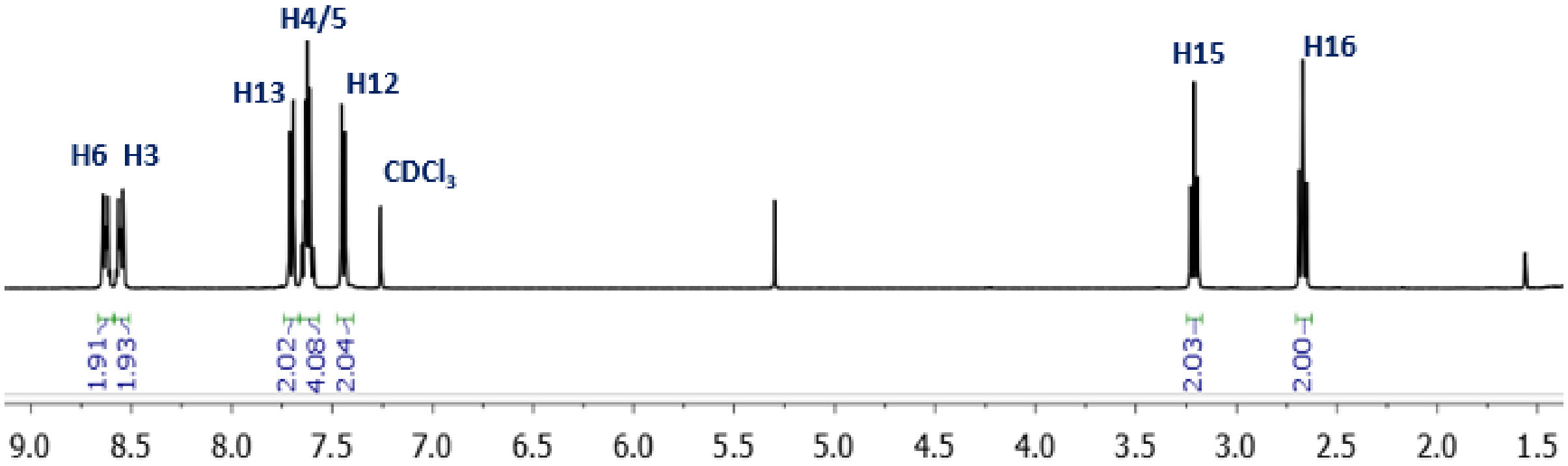

**Figure S1**: The $^1$H NMR spectrum of **5A** in $CDCl_3$.

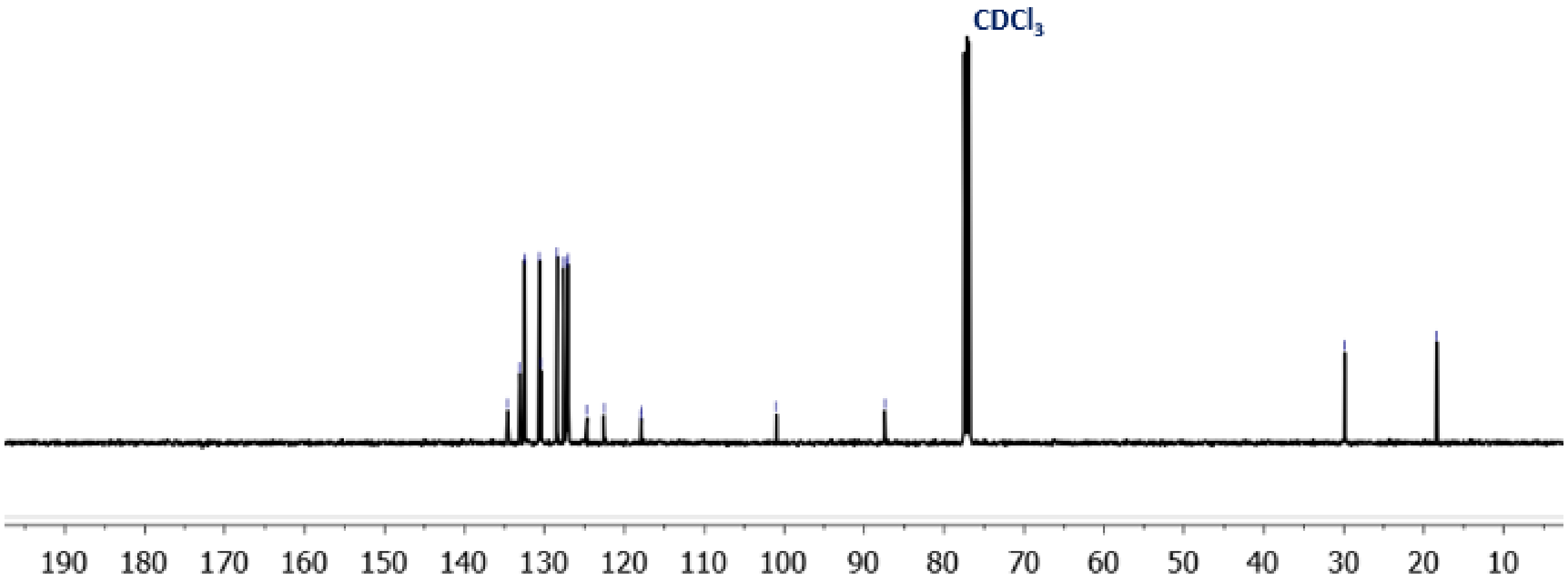

**Figure S2**: The $^{13}$C{$^1$H} NMR spectrum of **5A** in $CDCl_3$.

**9,10-Di(4-(ethynyl)phenylthiocyanoethyl)anthracene (8B)**

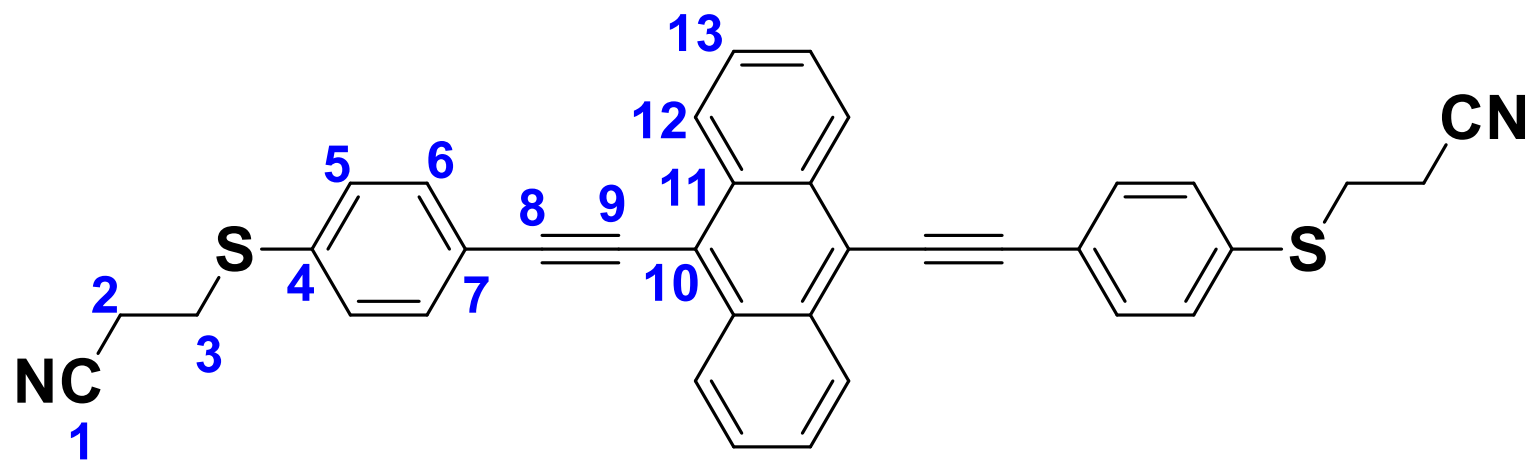

9,10-Dibromoanthracene (3.60 g, 10.72 mmol), 4-(ethynyl)phenylthiocyanoethyl (0.51 g, 2.71 mmol) and CuI (0.06 g, 0.29 mmol) were dissolved in DIPA (10 mL) and toluene (250 mL). $Pd(PPh_3)_2Cl_2$ (0.02 g, 0.33 mmol) was added and the mixture was stirred at 50°C for >16 hours to give a dark green solution. The solvent was removed *in vacuo* before the crude product was dissolved in DCM, filtered and subsequently purified by chromatography on an alumina V column, eluting with n-hexane/ethyl acetate (1:0 → 0:1). This compound was further purified through recrystallization from ethyl acetate to give a bright orange solid (0.26 g, 0.49 mmol, 18%, <5% impurity).

**$^1$H NMR** ($CDCl_3$, 298 K, 400 MHz): **$\delta_H$** = 8.69-8.64 (m, 4H, *H*12), 7.74 (dt, $^3J_{H\text{-}H}$ = 8.8, $^4J_{H\text{-}H}$ = 2.0 Hz, 2H, *H*5), 7.68-7.63 (m, 4H, *H*13), 7.47 (dt, $^3J_{H\text{-}H}$ = 8.8, $^4J_{H\text{-}H}$ = 2.0 Hz, 2H, *H*6), 3.21 (t, $^3J_{H\text{-}H}$ = 7.6 Hz, 2H, *H*3), 2.67 (t, $^3J_{H\text{-}H}$ = 7.6 Hz, 2H, *H*2) ppm; **$^{13}$C{$^1$H} NMR** ($CDCl_3$, 298 K, 100 MHz): **$\delta_C$** = 134.6 (Ar-C̲-C), 132.6 (Ar-C̲-H), 132.2 (Ar-C̲-C), 130.6 (Ar-C̲-H), 127.3 (Ar-C̲-H), 127.1 (Ar-C̲-H), 122.6 (Ar-C̲-C), 118.5 (Ar-C̲-C), 117.9 (Ar-C̲-C), 101.8 (-C̲≡C-), 87.9 (-C̲≡C-), 29.9 (-C̲$H_2$-), 18.4 (-C̲$H_2$-) ppm; **MS APCI**: calcd. [M]+ 549.1454; found. 549.1474.

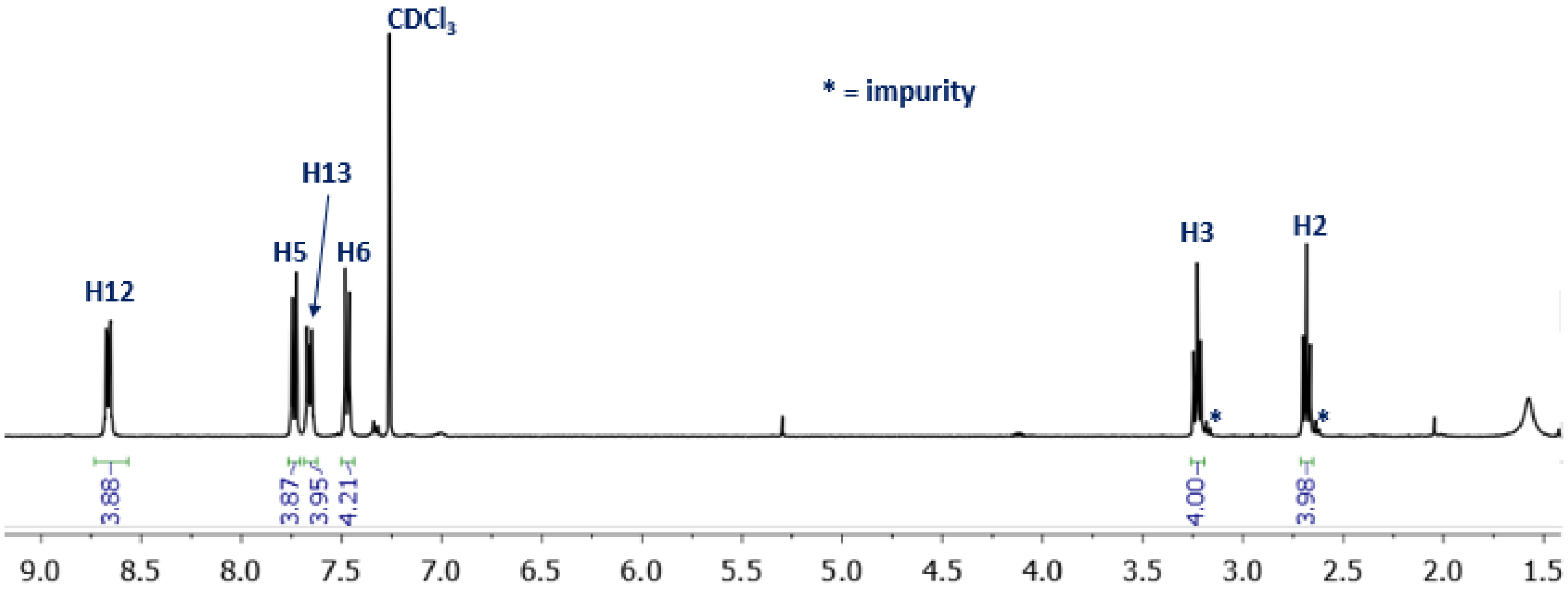

**Figure S3**: The $^1$H NMR spectrum of **5B** in $CDCl_3$.

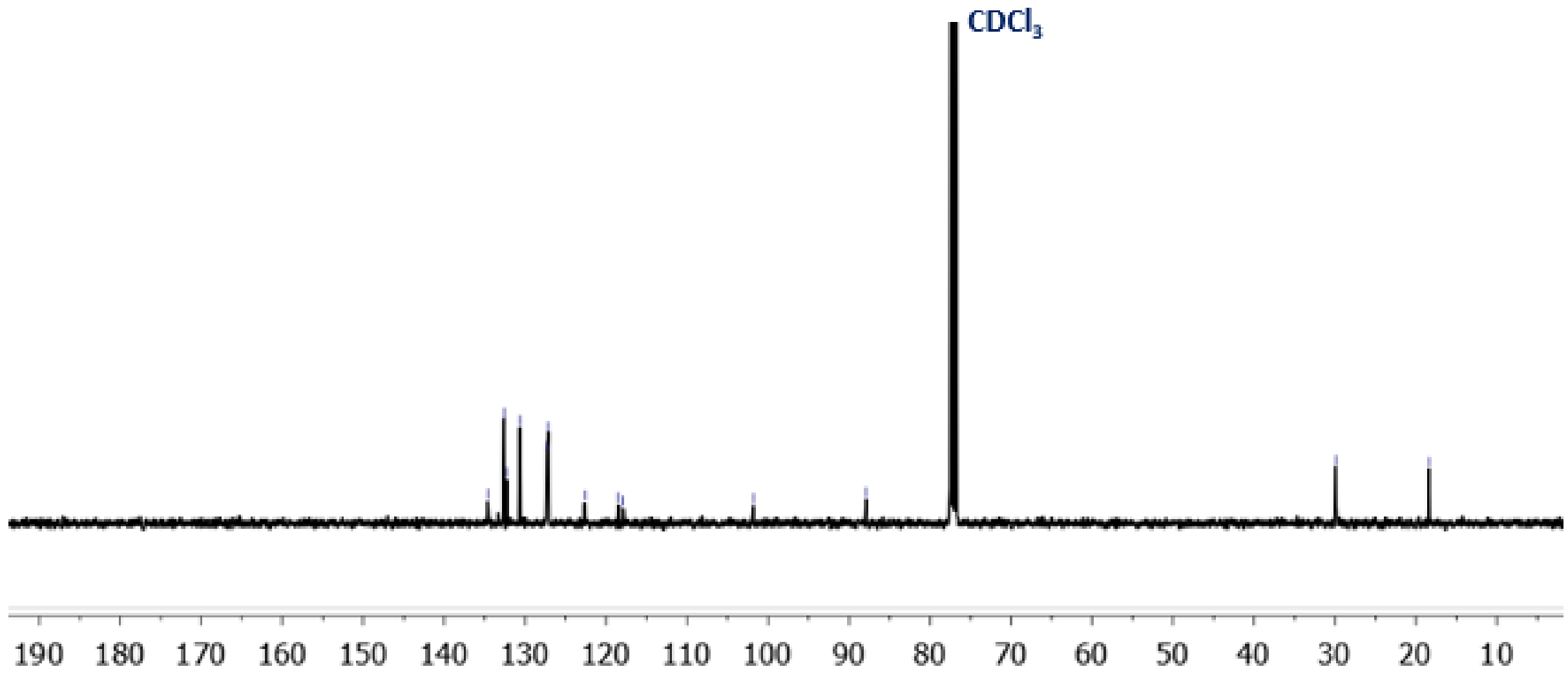

**Figure S4**: The $^{13}C\{^1H\}$ NMR spectrum of **5B** in $CDCl_3$.

**9-(4-(Ethynyl)pyridine)-10-(4-(ethynyl)phenylthiocyanoethyl)anthracene (5C)**

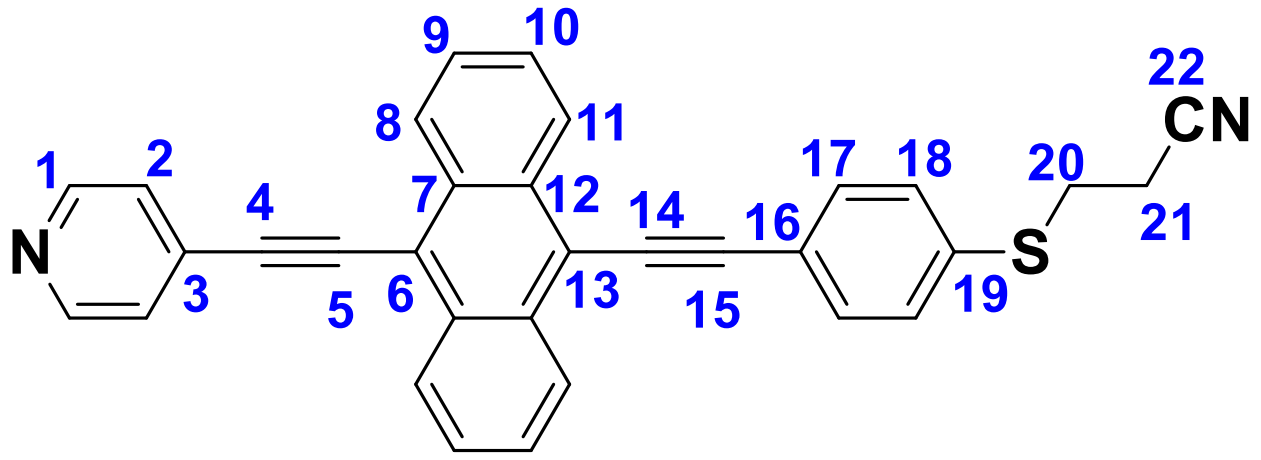

**5A** (0.26 g, 0.59 mmol), 4-(ethynyl)pyridine hydrochloride (0.18 g, 1.26 mmol) and CuI (0.04 g, 0.43 mmol) were dissolved in DIPA (10 mL) and DMF (150 mL). $Pd(PPh_3)_2Cl_2$ (0.05 g, 0.06 mmol) was added and the mixture was stirred at 75°C for >16 hours to give a dark red solution. The solution was diluted with ethyl acetate (300 mL) and washed with water (3 x 150 mL) before being dried over $MgSO_4$ and filtered. The solvent was removed *in vacuo* to give a bright orange solid that was purified by chromatography on an alumina V column, eluting with n-hexane/DCM (1:1 → 0:1), before being washed with copious amounts of hexane to give the final product as a bright orange solid (0.07 g, 0.15 mmol, 26%).

**$^1$H NMR** ($CDCl_3$, 298 K, 400 MHz): **$\delta_H$** = 8.72 (br s, 2H, *H*1), 8.68-8.59 (m, 4H, *H*8, *H*11), 7.73 (dt, $^3J_{H-H}$ = 8.8, $^4J_{H-H}$ = 2.0 Hz, 2H, *H*18), 7.70-7.63 (m, 4H, *H*9, *H*10), 7.62 (d, $^3J_{H-H}$ = 5.2 Hz, 2H, *H*2), 7.47 (dt, $^3J_{H-H}$ = 8.8, $^4J_{H-H}$ = 2.0 Hz, 2H, *H*17), 3.23 (t, $^3J_{H-H}$ = 7.6 Hz, 2H, *H*20), 2.68 (t, $^3J_{H-H}$ = 7.6 Hz, 2H, *H*21) ppm; **$^{13}C\{^1H\}$ NMR** ($CDCl_3$, 298 K, 100 MHz): **$\delta_C$** = 150.1 (Ar-<u>C</u>-H), 134.8 (Ar-<u>C</u>-C), 132.6 (Ar-<u>C</u>-H), 132.5 (Ar-<u>C</u>-C), 132.1 (Ar-<u>C</u>-C), 131.5 (Ar-<u>C</u>-C), 130.6 (Ar-<u>C</u>-H), 127.5 (Ar-<u>C</u>-H), 127.4 (Ar-<u>C</u>-H), 127.2 (Ar-<u>C</u>-H), 127.0 (Ar-<u>C</u>-H), 125.6 (Ar-

C-H), 122.4 (Ar-C-C), 119.6 (Ar-C-C), 117.9 (Ar-C-C), 117.2 (Ar-C-C), 102.3 (-C≡C-), 99.5 (-C≡C-), 91.0 (-C≡C-), 87.7 (-C≡C-), 29.9 (-$CH_2$-), 18.4 (-$CH_2$-) ppm; **MS APCI**: calcd. [M]+ 465.1420; found. 465.1420.

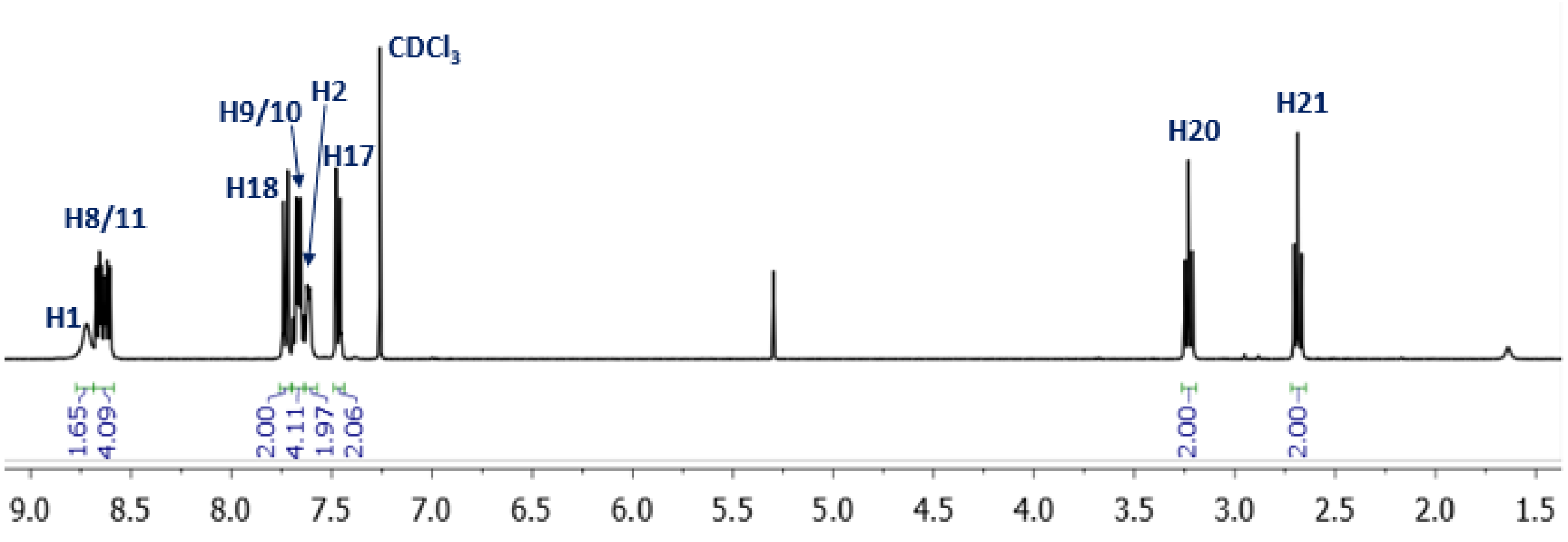

**Figure S5**: The $^{1}$H NMR spectrum of **5C** in $CDCl_3$.

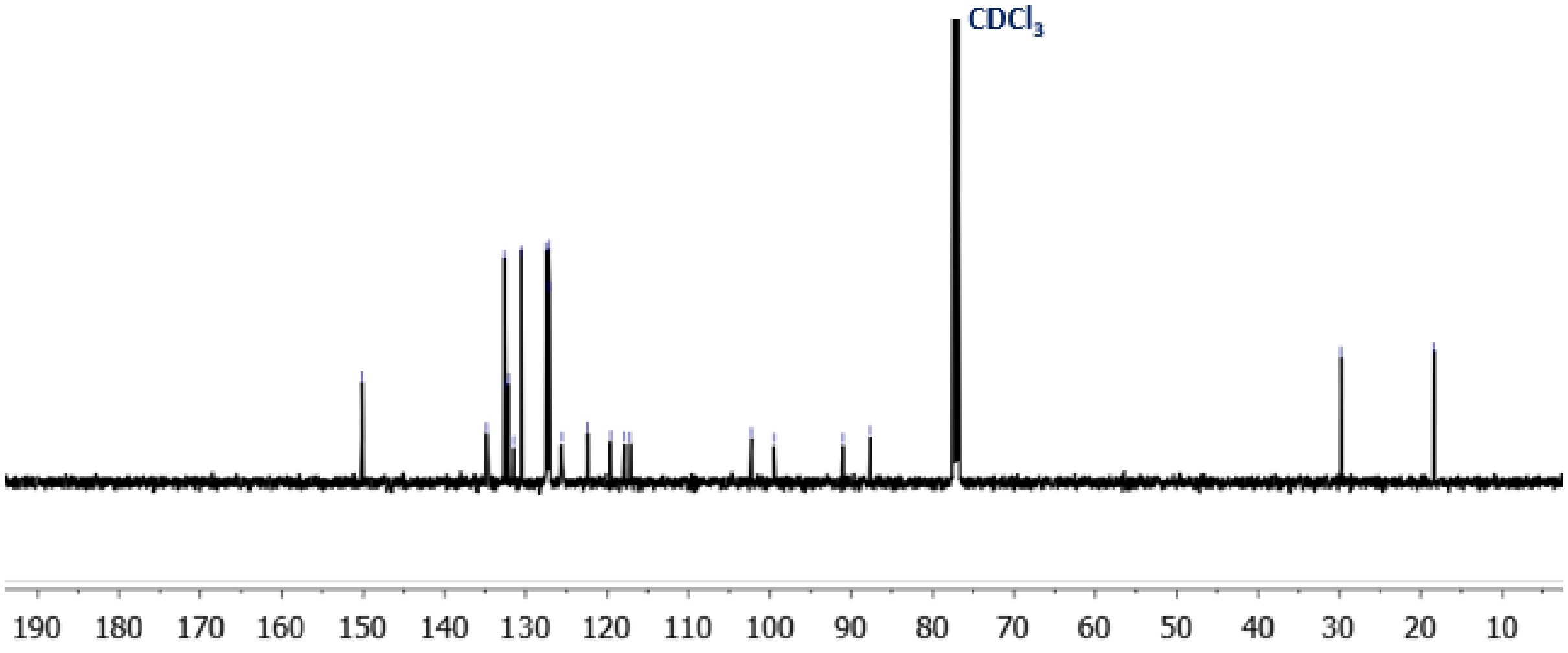

**Figure S6**: The $^{13}C\{^{1}H\}$ NMR spectrum of **8c** in $CDCl_3$.

**9-(4-(Ethynyl)pyridine)-10-(4-(ethynyl)phenylthioacetate)anthracene (5)**

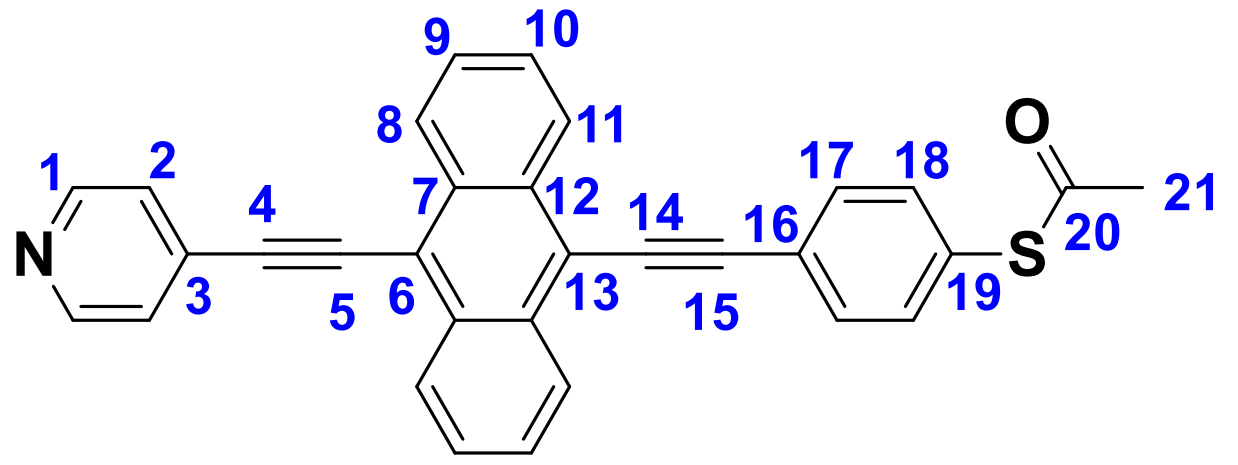

**5C** (0.06 g, 0.13 mmol) and NaOMe (0.25 g, 0.02 mmol) were dissolved in THF (20 mL) and stirred at room temperature for 1 hour to give a dark red solution. Acetic anhydride (1 M in THF, 0.50 mL) was added dropwise and the solution was left to stir for 30 minutes. The solvent was removed *in vacuo* and the crude product was dissolved in toluene (300 mL), washed with water (3 x 100 mL) and brine (100 mL),dried over $MgSO_4$ and filtered. The solvent was removed *in vacuo* to give the product as a bright orange solid (0.03 g, 0.07 mmol, 51%).

**$^{1}$H NMR** ($CDCl_3$, 298 K, 400 MHz): **$\delta_H$** = 8.72 (d, $^{3}J_{H-H}$ = 5.2 Hz, *H*1), 8.70-8.60 (m, 4H, *H*8, *H*11), 7.81 (dt, $^{3}J_{H-H}$ = 8.8, $^{4}J_{H-H}$ = 2.0 Hz, 2H, *H*18), 7.72-7.63 (m, 4H, *H*9, *H*10), 7.62 (dd, $^{3}J_{H-H}$ = 5.2, $^{4}J_{H-H}$ = 1.6 Hz, *H*2), 7.51 (dt, $^{3}J_{H-H}$ = 8.8, $^{4}J_{H-H}$ = 2.0 Hz, 2H, *H*17), 2.48 (s, 3H, *H*21) ppm; **$^{13}$C{$^{1}$H} NMR** ($CDCl_3$, 298 K, 100 MHz): **$\delta_C$** = 193.5 (-C=O), 150.1 (Ar-C-H), 134.6 (Ar-C-H), 132.5 (Ar-C-C), 132.4 (Ar-C-H), 132.2 (Ar-C-C), 131.5 (Ar-C-C), 129.0 (Ar-C-C), 127.5 (Ar-C-H), 127.5 (Ar-C-H), 127.3 (Ar-C-H), 127.0 (Ar-C-H), 125.6 (Ar-C-H), 124.5 (Ar-C-C), 119.5 (Ar-C-C), 117.3 (Ar-C-C), 102.3 (-C≡C-), 99.5 (-C≡C-), 91.1 (-C≡C-), 88.0 (-C≡C-), 30.5 (-$CH_3$) ppm; **MS APCI**: calcd. [M]+ 454.1260; found. 454.1241.

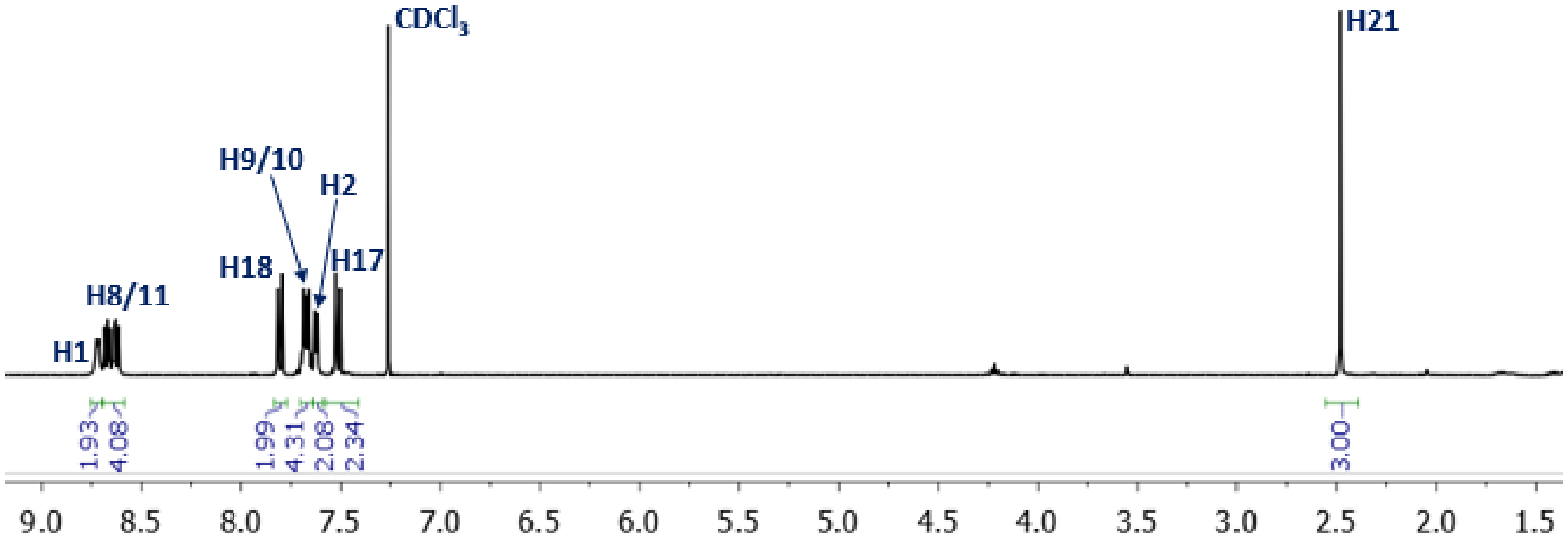

**Figure S7**: The $^{1}$H NMR spectrum of **5** in $CDCl_3$.

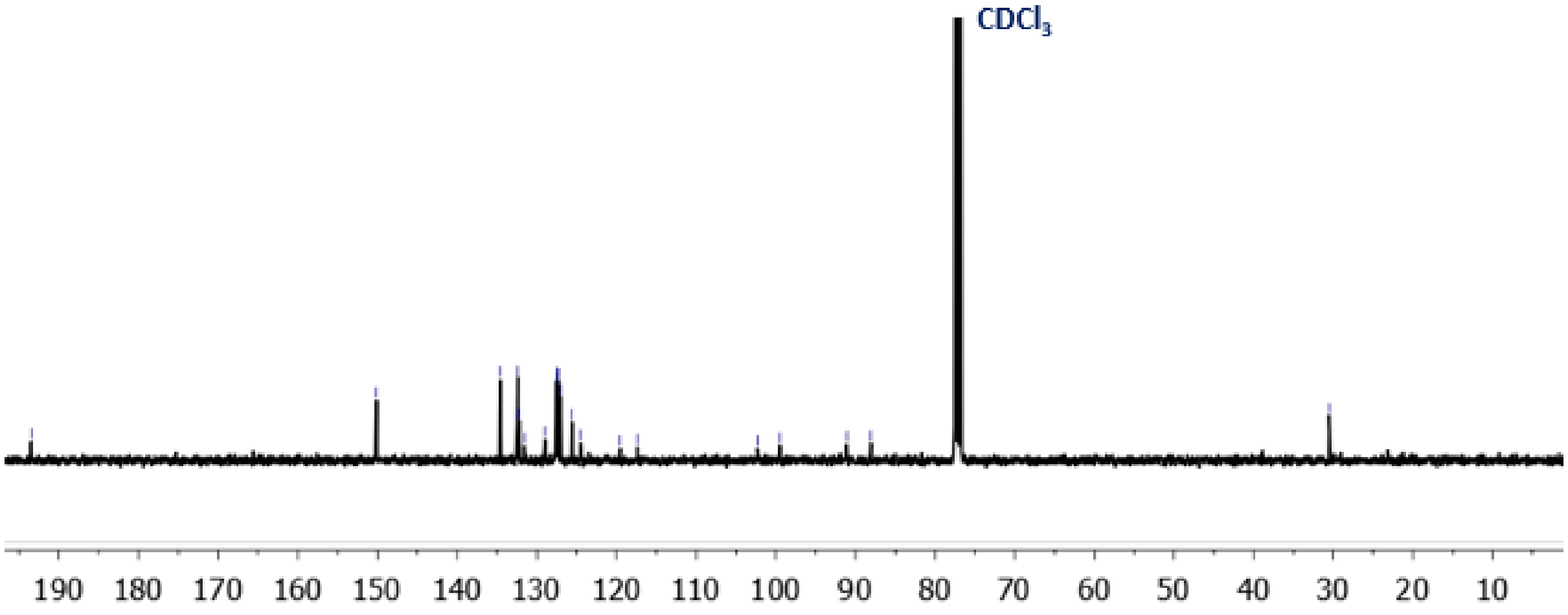

**Figure S8**: The $^{13}$C{$^{1}$H} NMR spectrum of **5** in $CDCl_3$.

## 2. Experimental Methods: tunnelling spectroscopy

2.1 STM BJ IV results for Au, and Molecules **1**, **4**, and **5**

As a measure of the background, without molecules present, Figs. S1 (a) –(c) show results obtained when no molecular adlayer was present on the Au substrate. Acquisition of data followed the same procedures, whereby a thermal gradient was applied and STM BJ IV measurements were performed obtaining results similar to Figs. 3(a) – (d) in the main text. Without a molecule present, sub-$G_0$ I/V sweeps were fewer in number, and occurred at shorter $\Delta z$. Nonetheless, a cluster similar to the green cluster in Figs. (c) and (d) yielded distributions of $\Delta V$, $G$, and $\Delta z$ values for different $\Delta T$. These were plotted in Figs. S1(a), (b), and (c), respectively. Linear fits for each variable across all $\Delta T$ were also plotted in black. The slope of $\Delta V$ vs $\Delta T$ determined a background Au/gap/Au thermopower, -0.2(2) μV $K^{-1}$, of our method and analysis (see Table 1). The slopes and standard errors on the slope for both $G$ and $\Delta z$ are summarized in Table S1. A complete discussion of dependences on thermal gradient can be found in section S2.5.

Complete results for molecule **2** are presented in section 2.2 and Fig. 4 of the main text. Results for molecules **1**, **4**, and **5** are plotted in Figs. S9(d)-(f) and S10. Replicate 1 (blue), 2 (red), and (when present) 3 (green) were plotted and fit to separate linear fits (navy, crimson, and dark green, respectively). As with molecule **5**, the single-molecule Seebeck coefficients, calculated from the slopes of the combined $\Delta V$ vs $\Delta T$ distributions, are plotted with black lines, as summarized in Table 1.

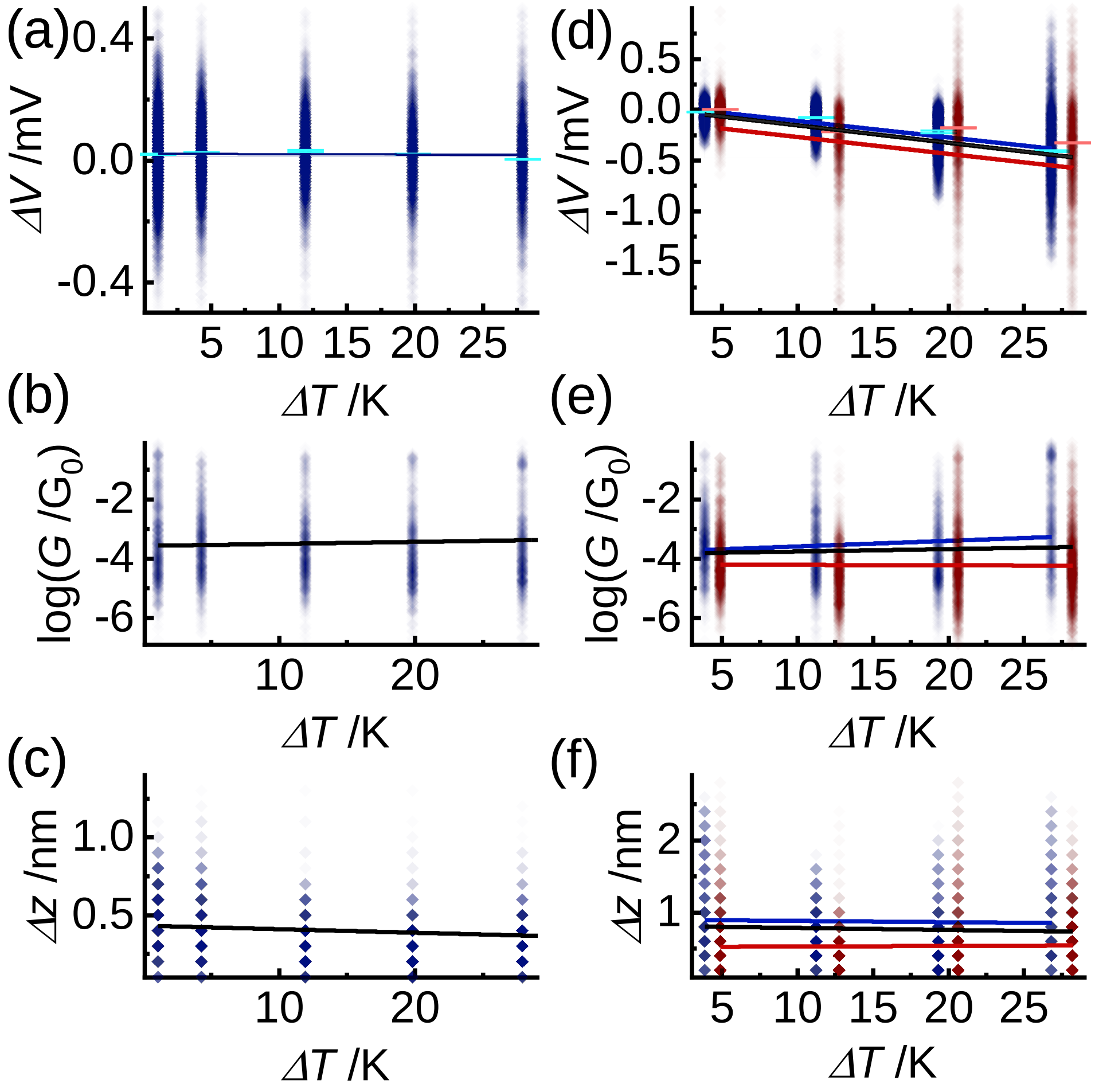

FIG S9. Scatter plots of *ΔV* vs *ΔT* measurements of (a) **Au/Au** with trend line in black, and (d) **1** with separate trend lines for two separate experiments (red and blue), and combined trend lines with 95% confidence intervals (black). Scatter plots of *G* vs *ΔT* measurements of (b) **Au/Au** with trend line in black, and (e) molecule **1** with separate trend lines for two separate experiments (red and blue), and combined trend lines with 95% confidence intervals (black). Scatter plots of *Δz* vs *ΔT* measurements of (c) **Au/Au** with trend line in black, and (f) molecule **1** with separate trend lines for two separate experiments (red and blue), and combined trend lines with 95% confidence intervals (black). Light blue and red cross hairs are mean and standard error of individual distributions, included as aids for the eye.

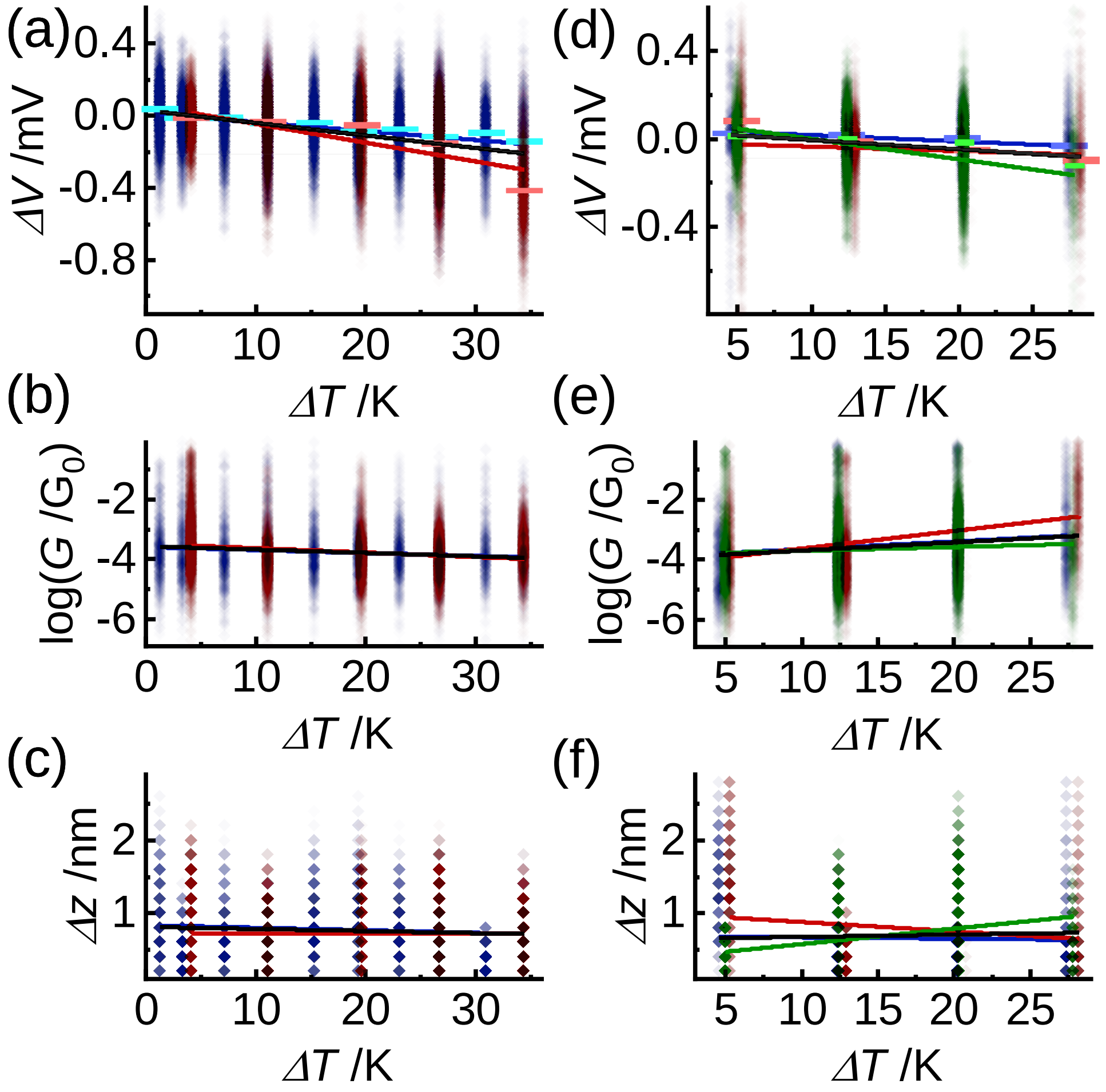

FIG S10. Scatter plots of *ΔV* vs *ΔT* measurements of (a) molecules **4** and (d) **5** with separate trend lines for two (for molecule **4**; three for molecule **5**) separate experiments (red and blue and green), and combined trend lines with 95% confidence intervals (black). Scatter plots of *G* vs *ΔT* measurements of (b) molecules **4** and (e) **5** with separate trend lines for two separate experiments (red and blue), and combined trend lines with 95% confidence intervals (black). Scatter plots of *Δz* vs *ΔT* measurements of (c) **4** and (f) **5** with separate trend lines for two separate experiments (red and blue), and combined trend lines with 95% confidence intervals (black). Light blue and red and green cross hairs are mean and standard error of individual distributions, included as aids for the eye.

2.2 Constant bias single-molecule break junctions

Constant bias results for molecules **1**, **3**-**5** are found in Figs. S11-S14. Constant bias results served to confirm the reliability of the molecule in the junction, and the obtain values for $G_{\mathrm{mol}}$ and $\Delta z_{mol}$ to validate the STM BJ IV results. Figs. S11-S14 were created following the same procedures outlined in section 2.3 in the main text. Notably, molecule **3** was successfully measured in STM BJ mode, but not in STM BJ IV mode, despite an intense effort. At present, the reason for this difference in behaviour is unclear, cf. main text.

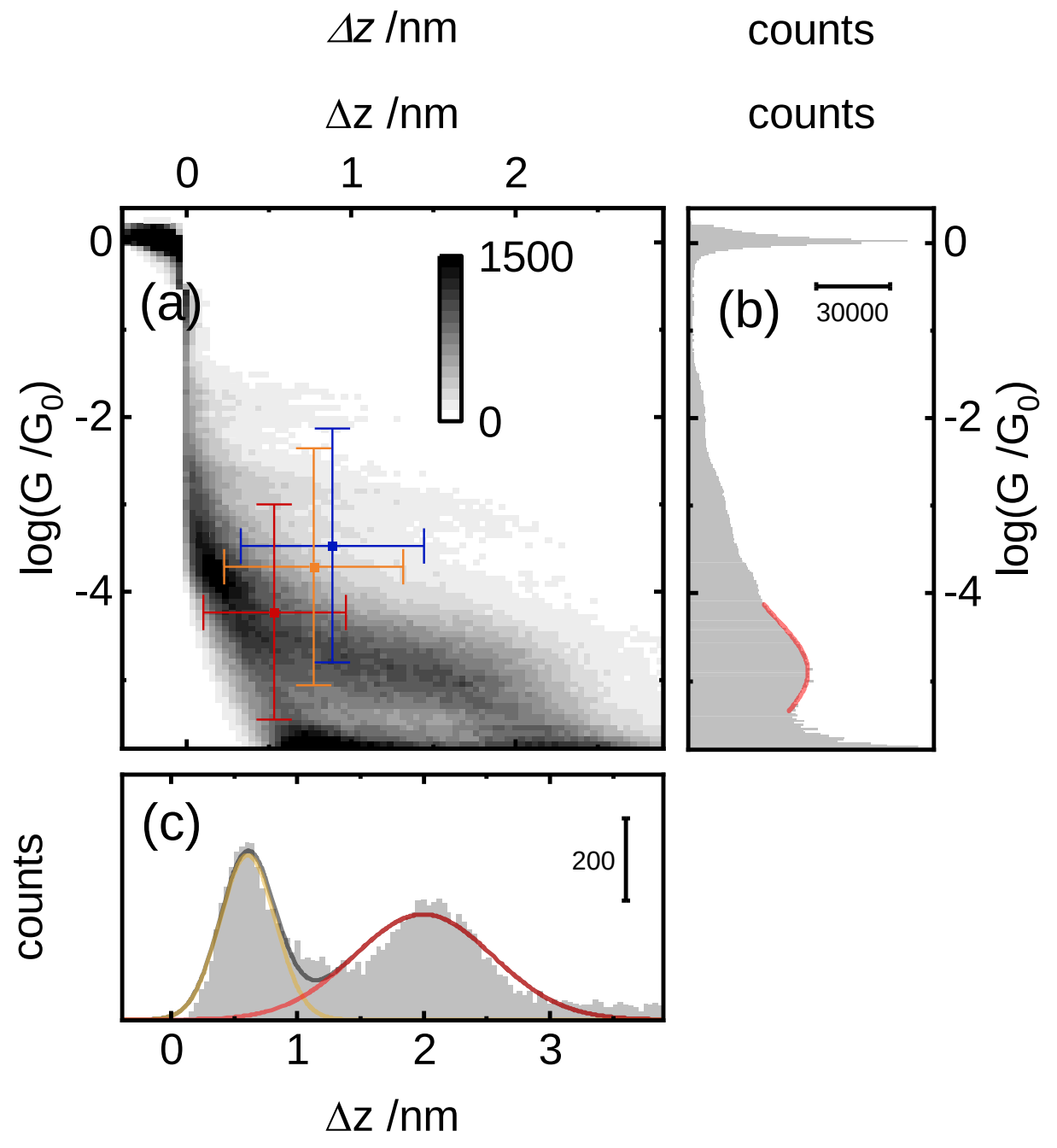

FIG S11. Example constant bias results for molecule **1** with c. 17k traces. (a) 2D conductance-displacement intensity plot of all traces. Cross hairs are trial 1 (blue) and trial 2 (red) and combined (orange) mean and standard deviation values of molecular cluster from STM BJ IV measurements. (b) 1D conductance histogram of all data in (a), with Gaussian peak fit (red). (c) Break-off distance histogram, determined at conductance $10^{-5.4}$ $G_0$, with two peak fit to separate tunnelling (gold) and molecular (red) trace displacements. Junction formation probability was 62%.

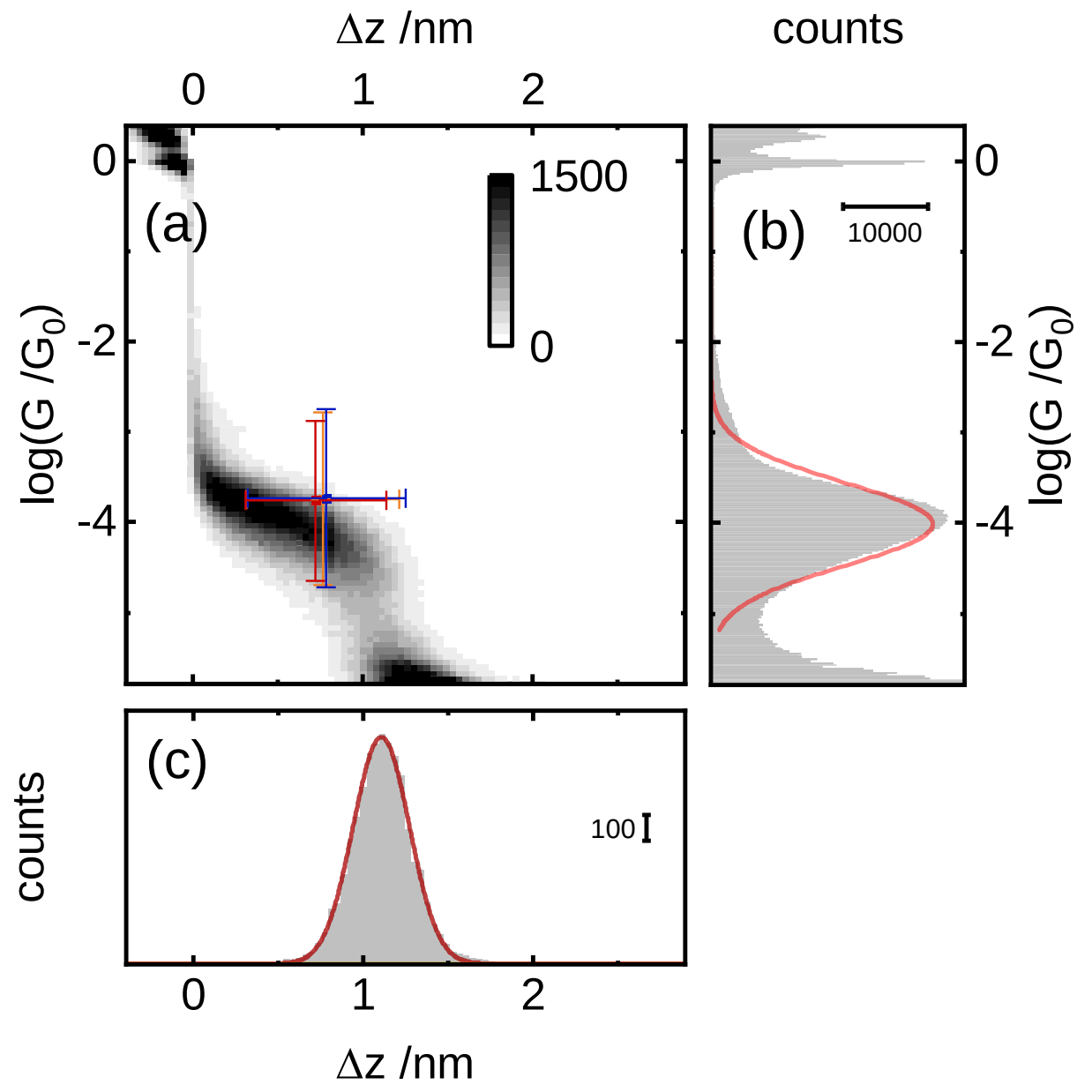

FIG S13. Example constant bias results for molecule **4**. (a) 2D conductance-displacement intensity plot of all traces. Cross hairs are trial 1 (blue), trial 2 (red), and combined (orange) mean and standard deviation values of molecular cluster from STM BJ IV measurements. (b) 1D conductance histogram of all data in (a), with Gaussian peak fit (red). (c) Break-off distance histogram, determined at conductance $10^{-5.2}$ $G_0$, with peak fit (red). Junction formation probability was close to 100%.

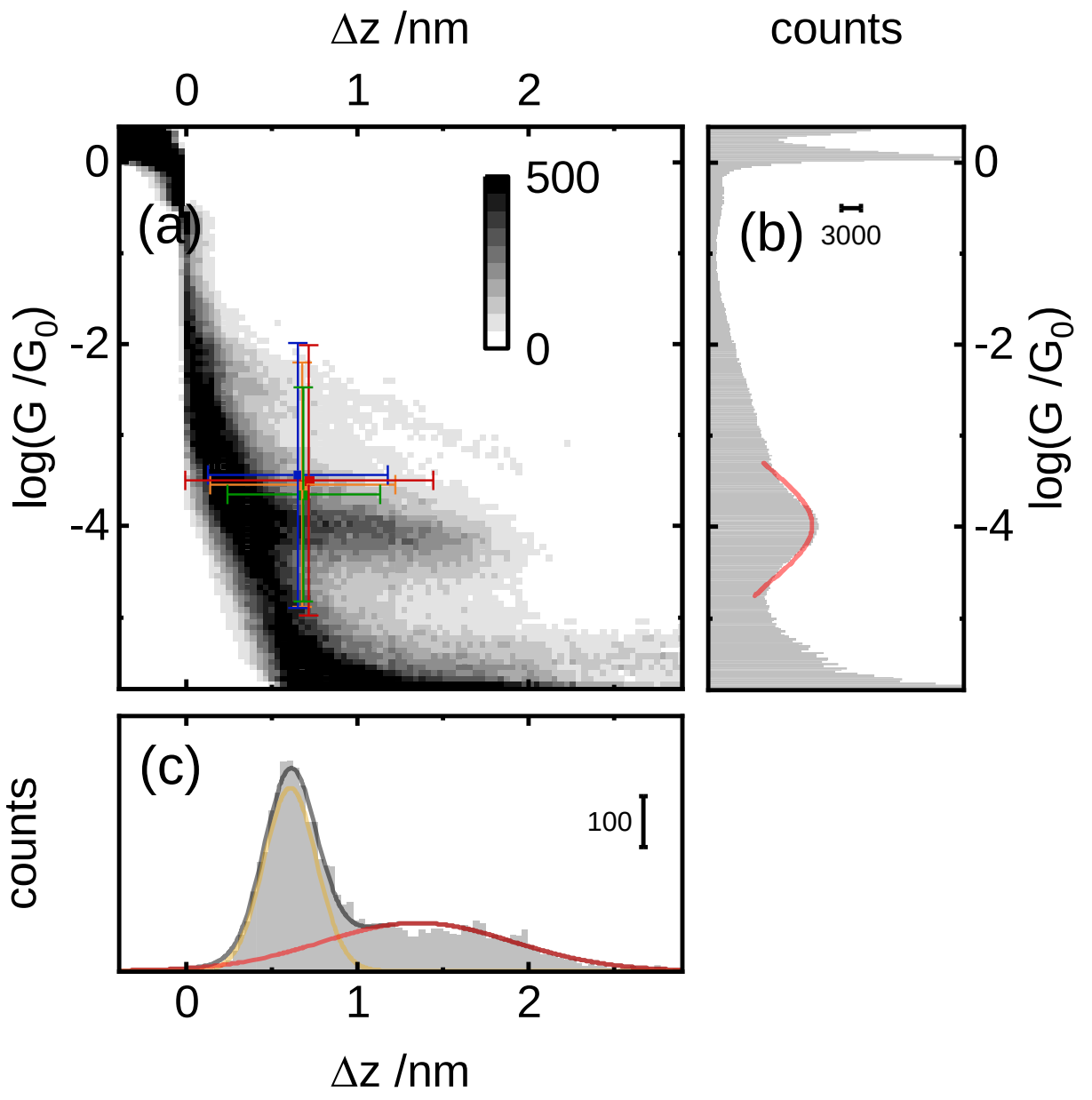

FIG S14. Example constant bias results for molecule **5**. (a) 2D conductance-displacement intensity plot of all traces. Cross hairs are trial 1 (blue), trial 2 (red), and trial 3 (green), and combined (orange) mean and standard deviation values of molecular cluster from STM BJ IV measurements. (b) 1D conductance histogram of all data in (a), with Gaussian peak fit (red). (c) Break-off distance histogram, determined at conductance $10^{-5.2}$ $G_0$, with 2 peak fit to separate tunnelling (yellow) and molecular (red) trace displacements. Junction formation probability was 49%.

2.3 Temperature dependence of $\Delta z_{\mathrm{mol}}$ and $G_{\mathrm{mol}}$

Table S1 summarizes the slopes and standard errors for *Δz* and *G*. The expected charge transport regime for all molecules measured in this study was coherent transport. As a consequence, we expect there to be no (significant) dependence of molecular conductance on the applied temperature gradient. This is indeed borne out in the data. Likewise, there does not seem to be a systematic and statistically significant dependence of the molecular break-off distance on DT, which might suggest that the junction stability is not affected the change in temperature gradient.

Table S1. Summary of temperature dependencies.

| | **Δz** | *error* | **G** | *error* |
|---|---|---|---|---|
| *Molecule* | *Slope* nm $K^{-1}$ | *Std Err* nm $K^{-1}$ | *Slope* $\log(G\,G_0^{-1})\,K^{-1}$ | *Std Err* $\log(G\,G_0^{-1})\,K^{-1}$ |
| **1** (1,5 SAc2) | -2.8E-3 | 4.1E-4 | 8.2E-3 | 1.0E-3 |
| **2** (9,10 SAc2) | 7.1E-3 | 3.8E-4 | 1.3E-2 | 6.5E-4 |
| ***3*** *(*9,10 SMe2*)* | --- | --- | --- | --- |
| **4** (9,10 SMe,N) | -2.8E-3 | 1.8E-4 | -1.1E-2 | 3.8E-4 |
| **5** (9,10 SAc,N) | 3.1E-3 | 4.7E-4 | 2.7E-2 | 1.1E-3 |
| ***Au*** | -2.2E-3 | 1.0E-4 | 6.8E-3 | 6.4E-2 |

## 3. Theoretical details

### 3.1 DFT and Transport Calculations

The ground state Hamiltonian and optimized geometry of each molecule was obtained using the density functional theory (DFT) code SIESTA.[5,6] The local density approximation (LDA) exchange correlation functional was used along with double zeta polarized (DZP) basis sets and the norm conserving pseudo potentials. The real space grid was defined by a plane wave cut-off of 250 Ry. Geometry optimization was carried out to a force tolerance of 0.01 eV/Å. This process was repeated for a unit cell with the molecule located between gold electrodes, in which the optimized distances between Au and the anchor groups were obtained. From the ground state Hamiltonian, the transmission coefficient, the room temperature electrical conductance $G$ and Seebeck coefficient $S$ were obtained using the GOLLUM quantum transport code.[7]

### 3.2 Optimised DFT Structures of Isolated Molecules

Fig S15 shows the optimum geometries of the isolated molecules **1-5**, obtained by using the local density approximation (LDA). We also computed results using GGA and found that the resulting transmission functions were comparable with those obtained using LDA.[8,9,10] Molecules **1-4** of the 5 molecules were studied previously.[2]

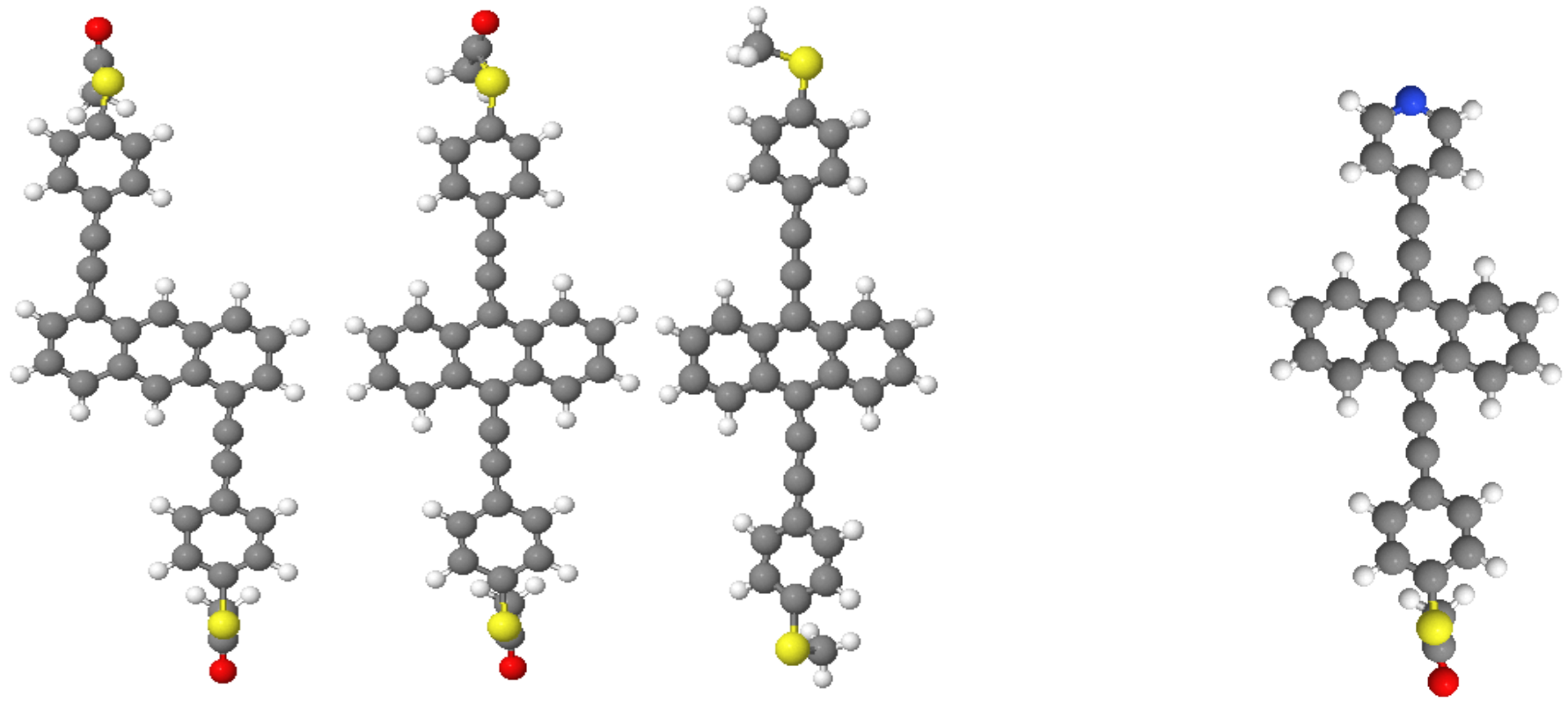

FIG S15. Fully relaxed isolated molecules **1-5** **(1-4** previously reported[2]**)**. Key: C = grey, H = white, O = red, S = yellow, N = blue.

### 3.3 Binding energy of molecules on Au

To calculate the optimum binding distance between N/S/SMe anchor groups and Au(111) surfaces, we used DFT and the counterpoise method, which removes basis set superposition errors (BSSE). The binding distance is defined as the distance between the gold surface and the N/S/SMe terminus of the anchor groups. Here, the molecule is defined to be entity A and the gold electrode to be entity B. The ground state energy of the total system is calculated using SIESTA and is denoted $E_{AB}^{AB}$. The energy of each entity is then calculated in a fixed basis, which is achieved using ghost atoms in SIESTA. Hence, the energy of A in the presence of the fixed basis is defined as $E_{A}^{AB}$ and for the gold as $E_{B}^{AB}$. The binding energy is then calculated using the following equation:

$$\text{Binding Energy} = E_{AB}^{AB} - E_{A}^{AB} - E_{B}^{AB} \qquad \text{(S1)}$$

We then considered the nature of the binding depending on the gold surface structure. We calculated the binding to a Au pyramid on a surface with the N/S/SMe atom/s binding at a 'top' site and then varied the binding distance. Figure S16 (left) shows that value of distances = 2.3, 2.4 and 2.7 Å give the optimum distances, at approximately 0.5, 0.8 and 0.4 eV (N/S/SMe respectively).

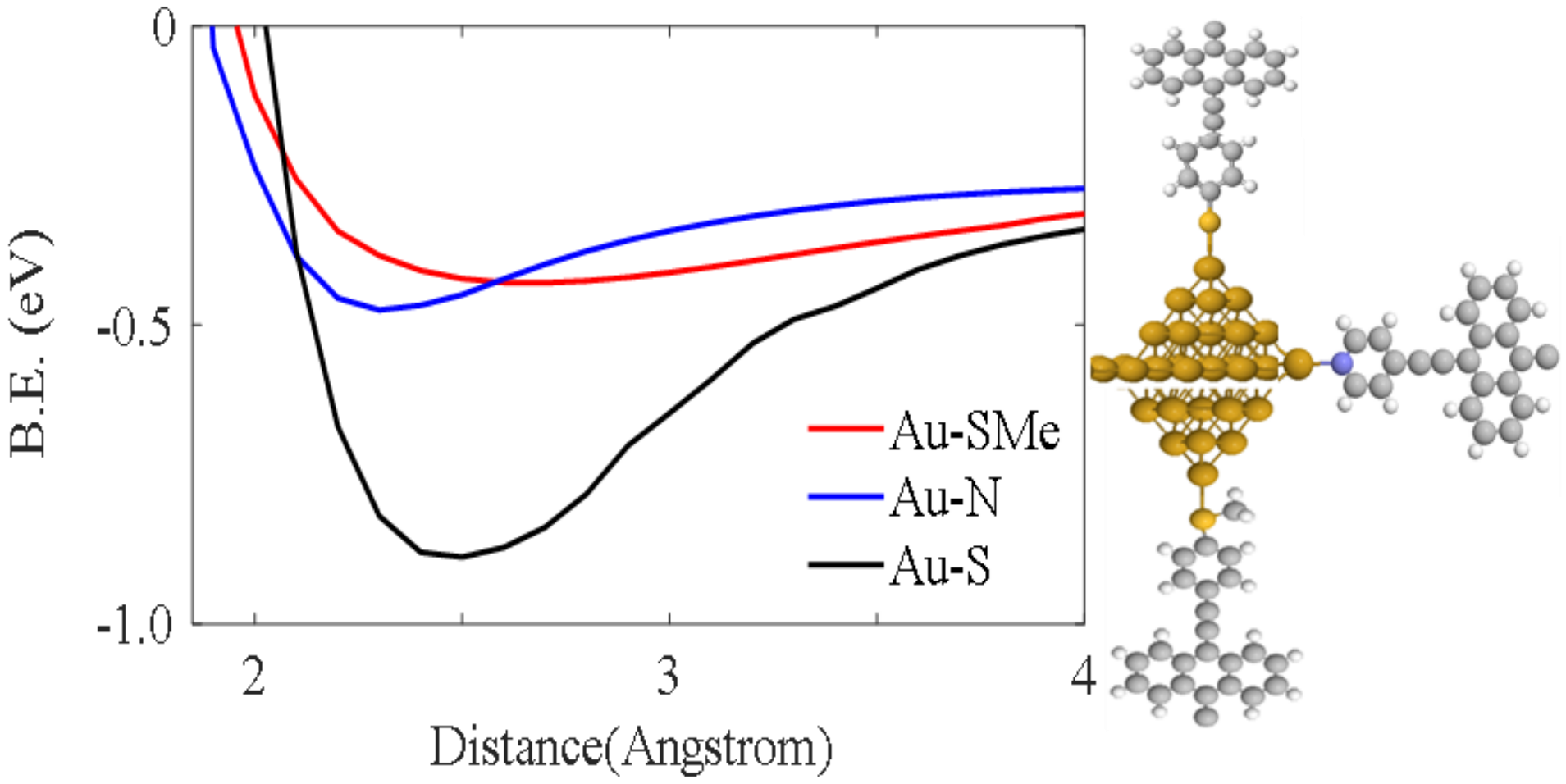

FIG S16. Example of a binding energy plot for three different anchors Au-SMe, Au-N and Au-S (left), with an idealised adatom configuration at the Au lead interface (right, Au-S, Au-N and Au-SMe). Key: C = grey, H = white, N = blue, S = light yellow, Au = dark yellow.

3.4 Optimised DFT Structures of Compounds within Junctions

Using the optimised structures and geometries for the isolated compounds obtained above, we again employed the SIESTA code to calculate self-consistent optimised geometries, ground state Hamiltonians and overlap matrix elements for each metal-molecule-metal junction. Leads were modelled as 625 atom slabs, terminated with 11-atom Au(111) tips. The optimised structures were then used to compute the transmission curve for each compound. The DFT optimised geometries **1-5** are shown here, in Figures S17-S21. Key: C = grey, H = white, S = light yellow, N = blue, Au = dark yellow.

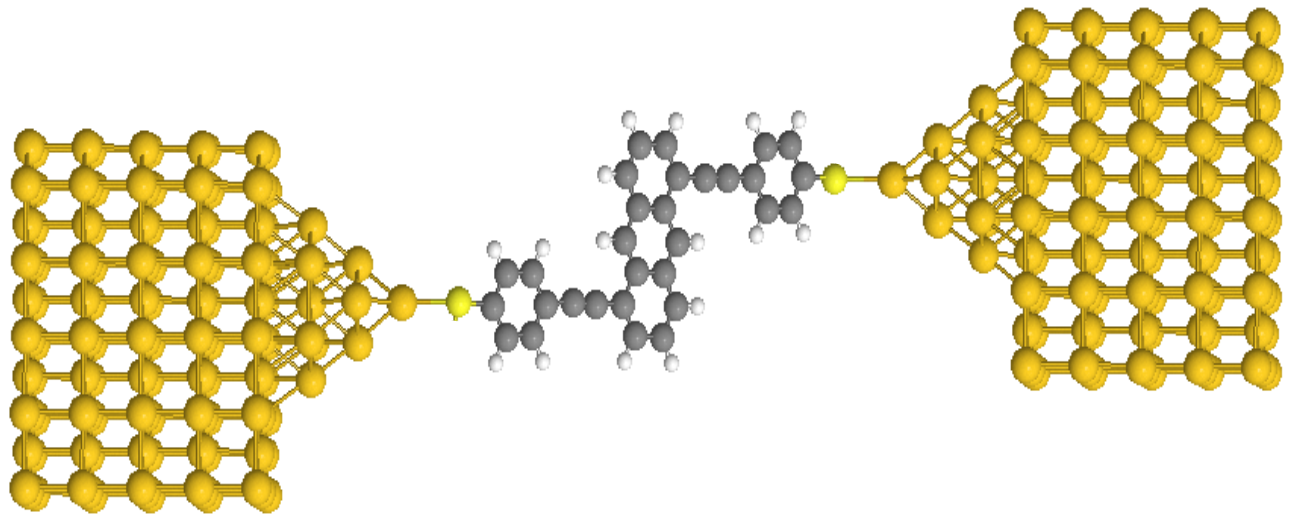

FIG S17. Optimised structure of **1,** (1,5-Di(4-(ethynylphenyl)thioacetate) anthracene).

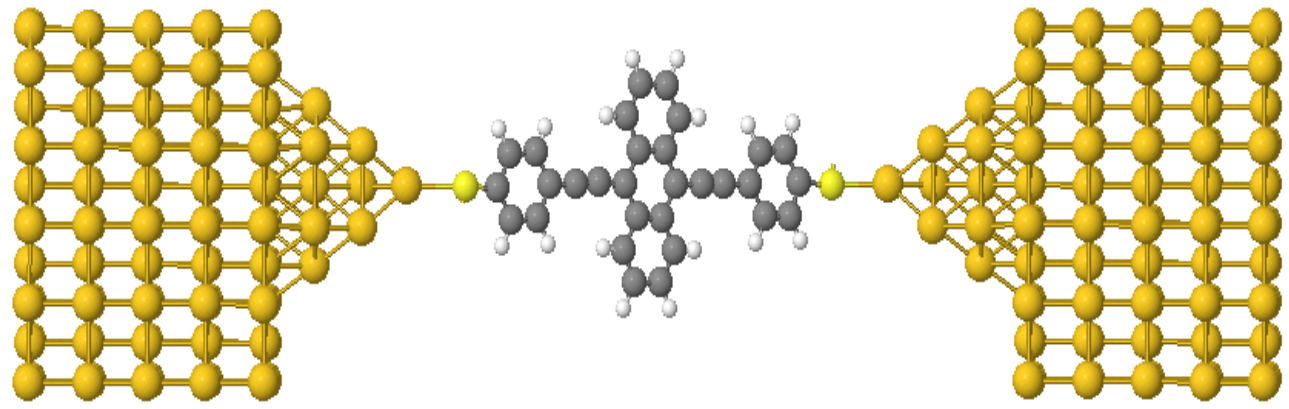

FIG S18. Optimised structure of **2**, (9,10-Di(4-(ethynylphenyl)thioacetate)anthracene).

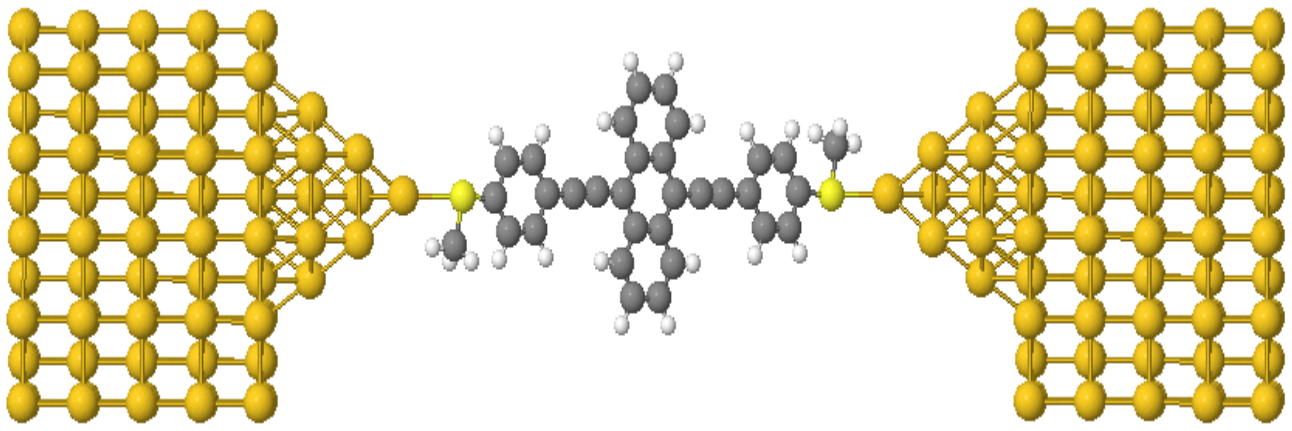

FIG S19. Optimised structure of **3**, (9,10-Di(4-ethynylthioanisole)anthracene).

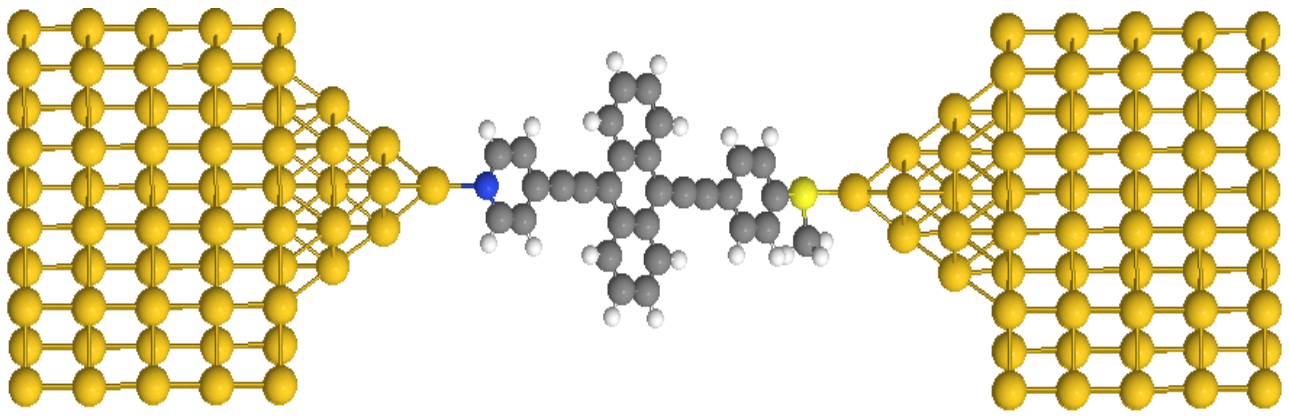

FIG S20. Optimised structure of **4**, (9-(4-(ethynylphenyl)thioacetate)-10-(4-ethynylpyridine)anthracene).

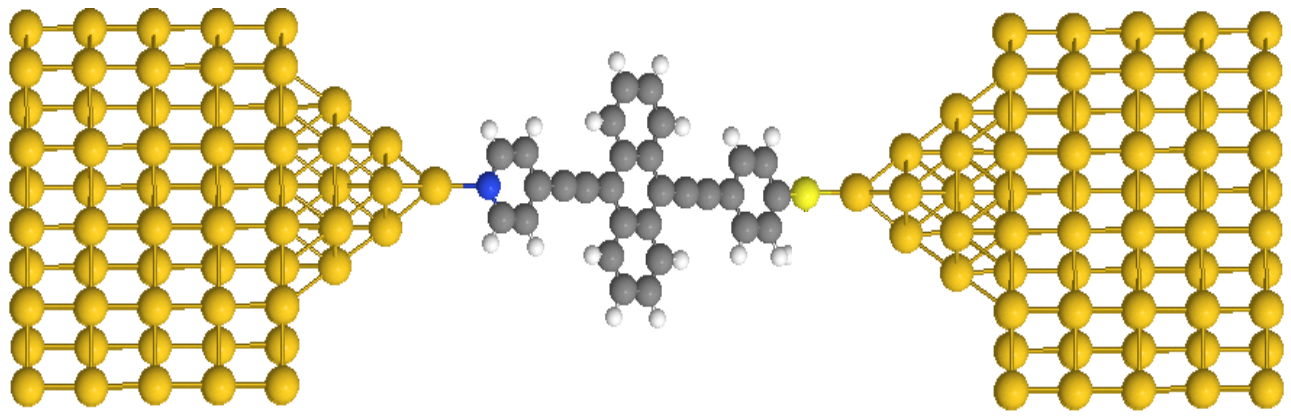

FIG S21. Optimised structure of **5**, (9-(4-ethynlthioanisole)-10-(4-ethynylpyridine)anthracene).

3.5 Transport Calculations

The transmission coefficient curves $T(E)$, obtained from using the GOLLUM transport code, were calculated for the five compounds **1-5.** Transmission coefficient curves for **1-4** were reported in Ref 2). The HOMO resonance is predicted to be pinned near the Fermi Level of the electrodes for molecules **1** and **2** whereas, the LUMO resonance is predicted to be pinned near the Fermi level of the electrodes for **3-5**. In practice however, we expect Fermi Level to be in the vicinity of the mid gap for each molecule (black-dashed line), as shown in Figures S22-S31.

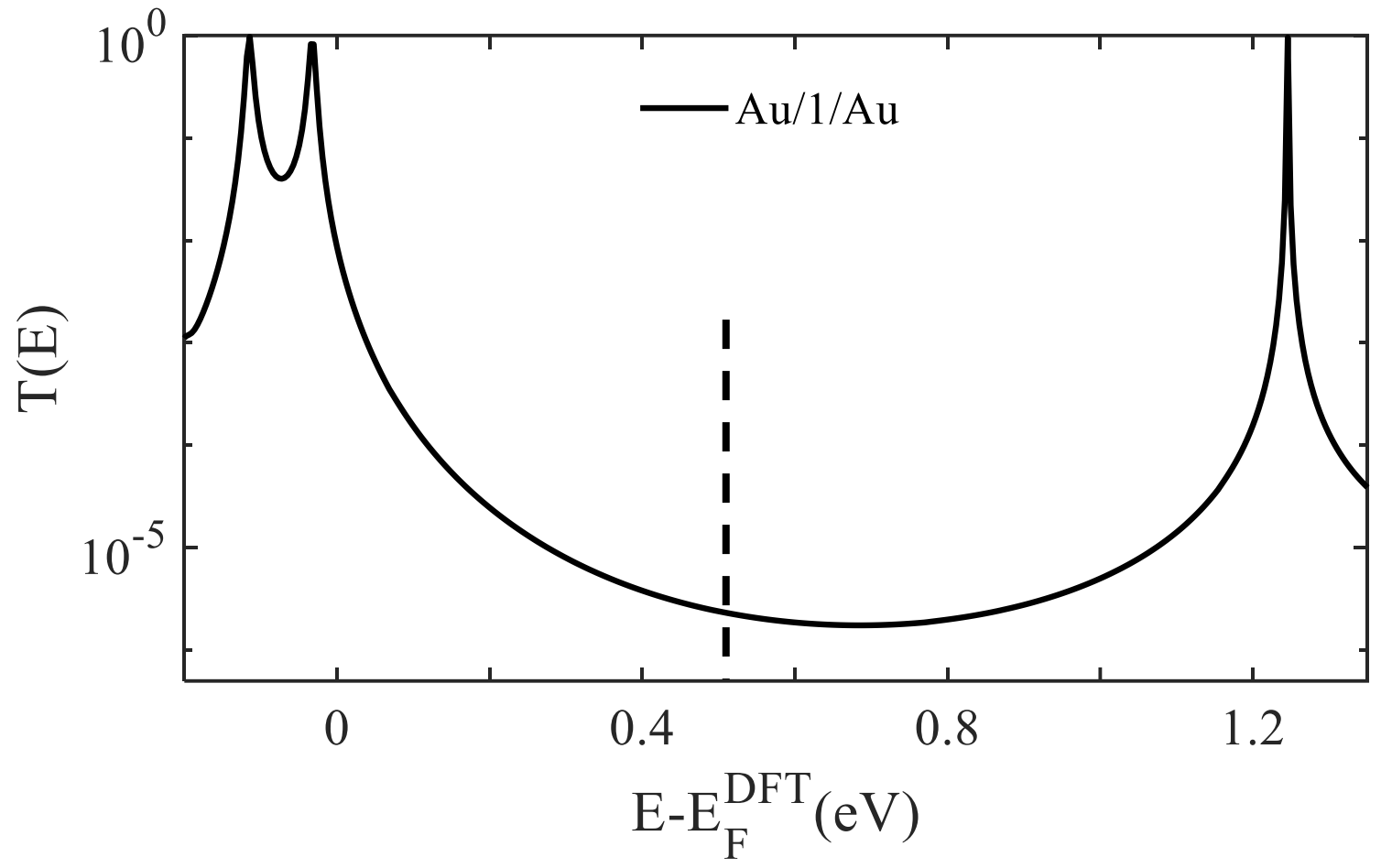

FIG S22. **Transmission coefficients of 1, (1,5-Di(4-(ethynylphenyl)thioacetate) anthracene.** Zero bias transmission coefficient T(E) **1,** (1,5-Di(4-(ethynylphenyl)thioacetate) anthracene, against electron energy E.

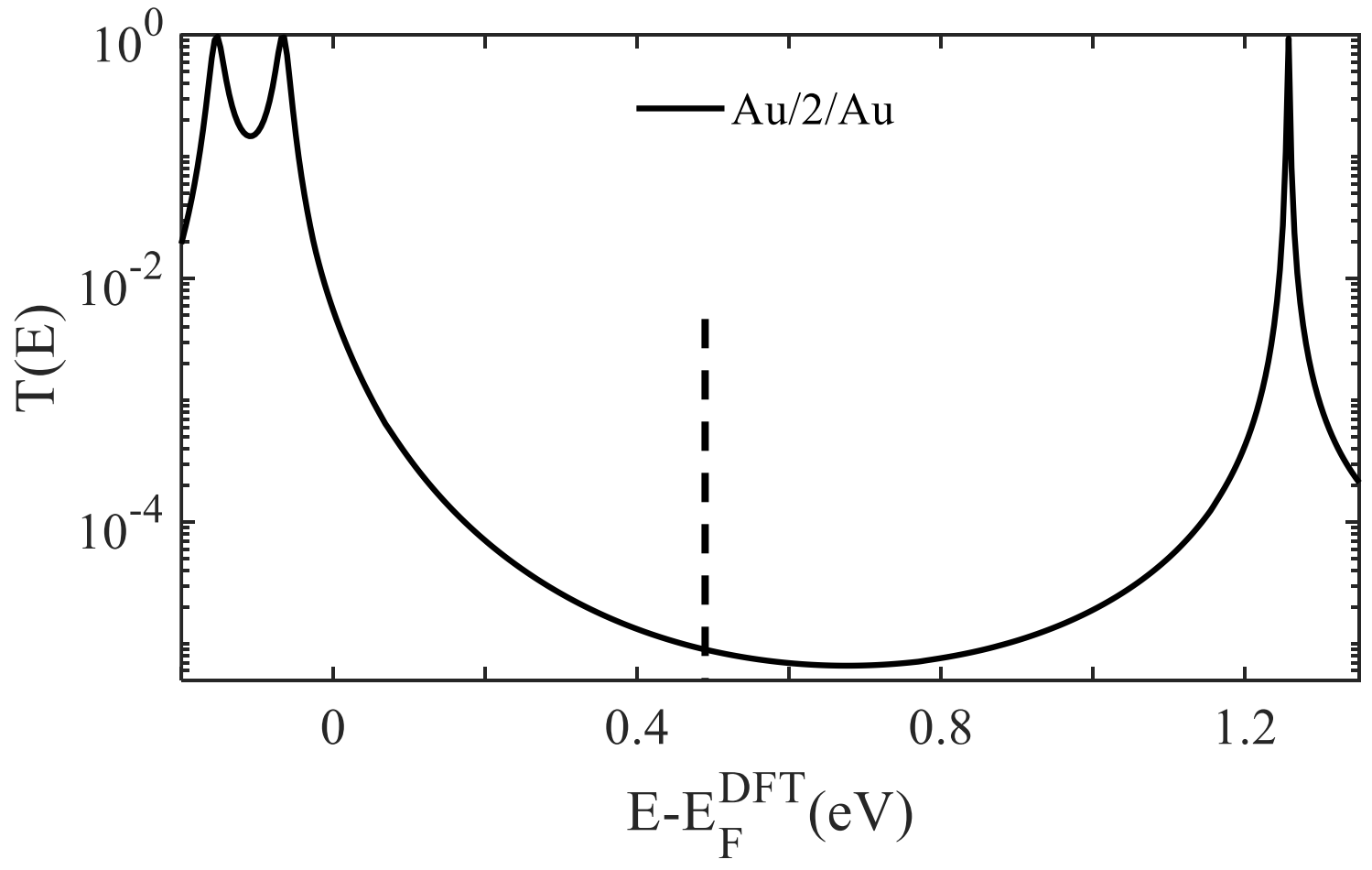

FIG S23. **Transmission coefficients of 2**, **(9,10-Di(4-(ethynylphenyl)thioacetate) anthracene.** Zero bias transmission coefficient T(E) of **2**, (9,10-Di(4-(ethynylphenyl)thioacetate) anthracene, against electron energy E.

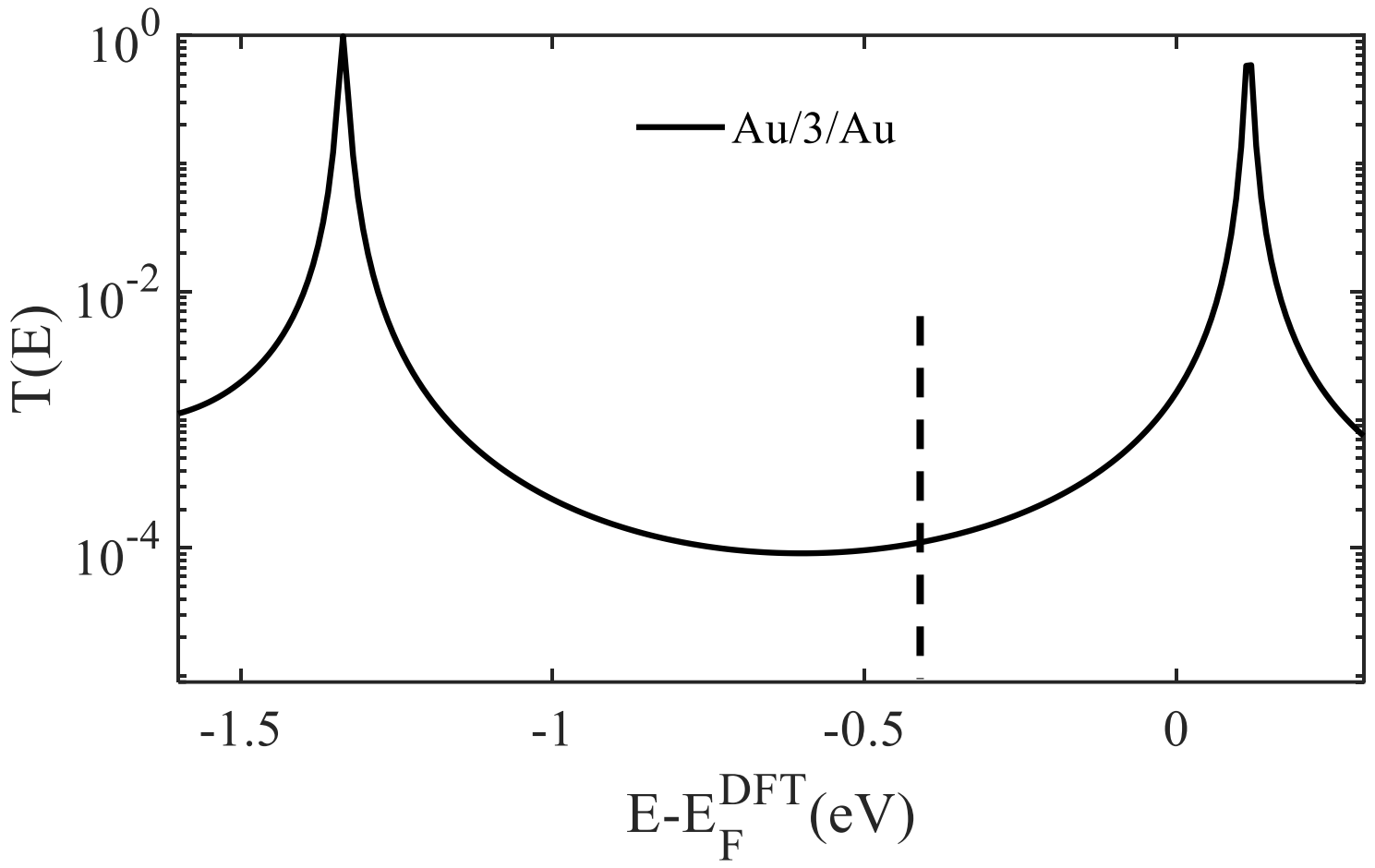

FIG S24. **Transmission coefficients of 3**, **(9,10-Di(4-ethynylthioanisole)anthracene**. Zero bias transmission coefficient T(E) of **3**, (9,10-Di(4-ethynylthioanisole)anthracene, against electron energy E.

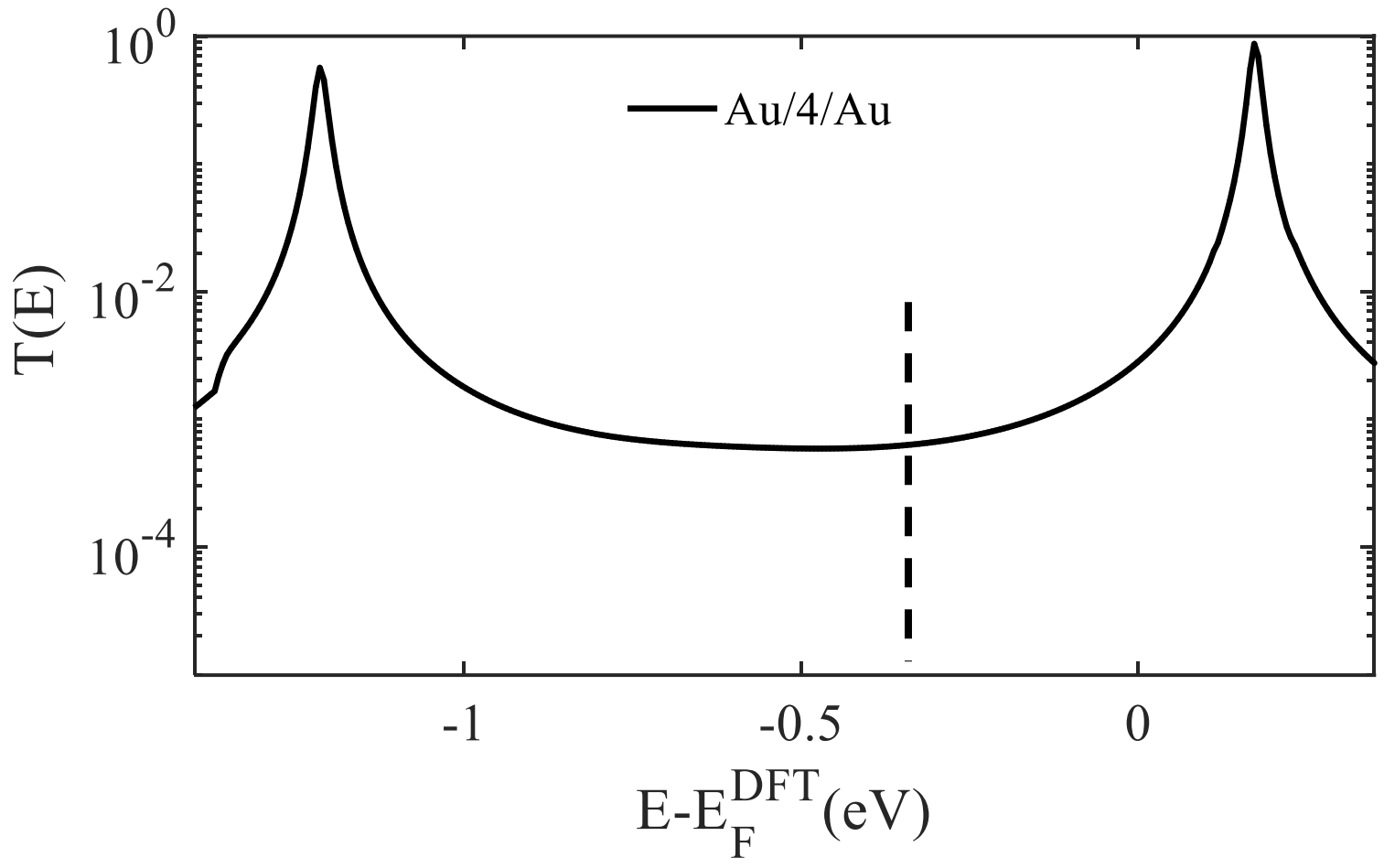

FIG S25. **Transmission coefficients of 4**, **(9-(4-(ethynylphenyl)thioacetate)-10-(4-ethynylpyridine) anthracene.** Zero bias transmission coefficient T(E) of **4**, (9-(4-(ethynylphenyl)thioacetate)-10-(4-ethynylpyridine)anthracene, against electron energy E.

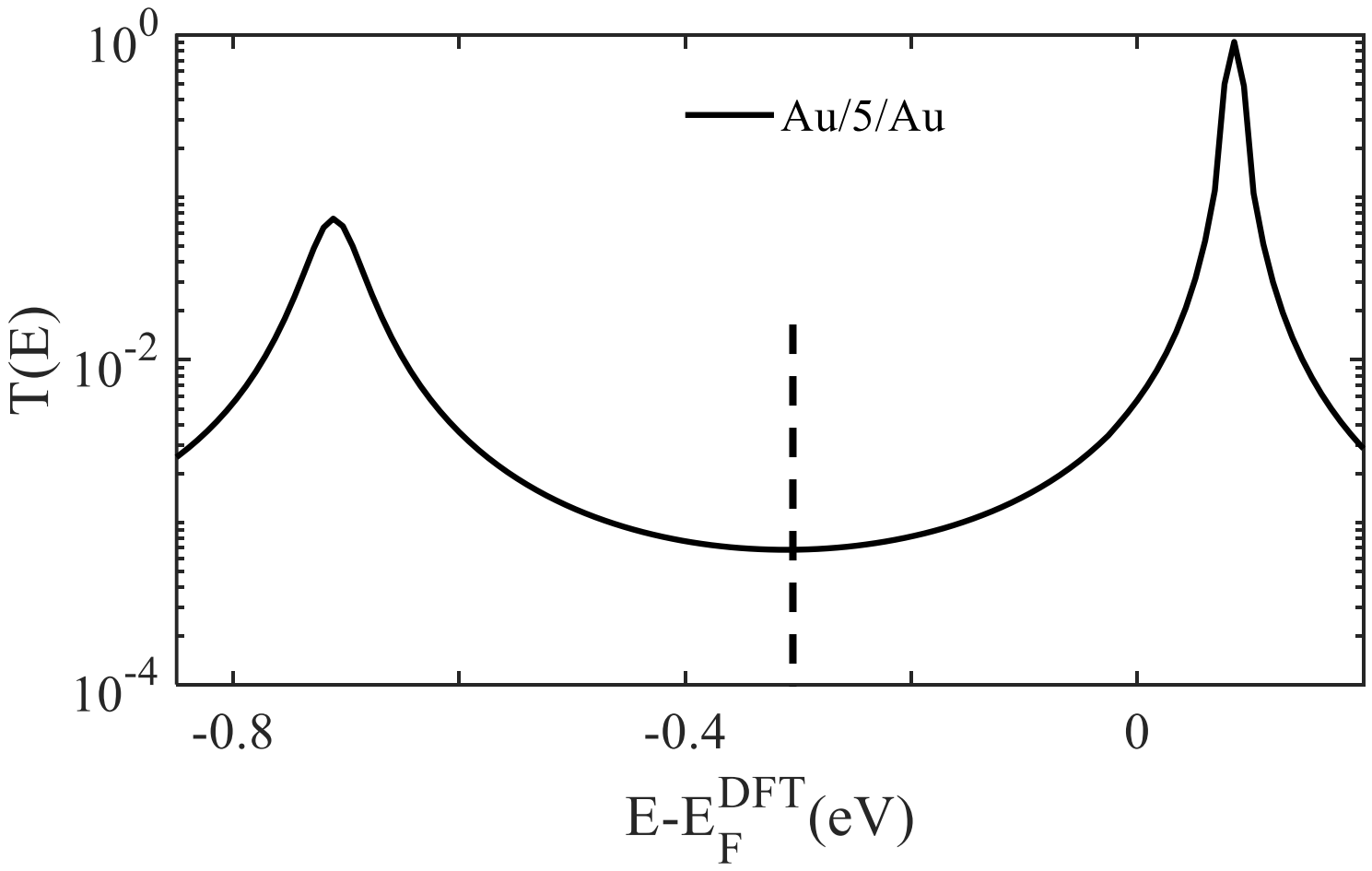

FIG S26. **Transmission coefficients of 5**, **(9-(4-ethynlthioanisole)-10-(4-ethynylpyridine)anthracene.** Zero bias transmission coefficient T(E) of **5**, (9-(4-ethynlthioanisole)-10-(4-ethynylpyridine)anthracene, against electron energy E.

3.6 Seebeck coefficients

To calculate the Seebeck cofficient of these molecular junctions, it is useful to introduce the non-normalised probability distribution $P(E)$ defined by

$$P(E) = -T(E)\frac{df(E)}{dE} \qquad \text{(S2)}$$

where $f(E)$ is the Fermi-Dirac function and $T(E)$ are the transmission coefficients and whose moments $L_i$ are denoted as follows

$$L_i = \int dEP(E)(E - E_F)^i \qquad \text{(S3)}$$

where $E_F$ is the Fermi energy. The Seebeck cofficient $S$, is then given by

$$S(T) = -\frac{1}{|e|T}\frac{L_1}{L_0} \qquad \text{(S4)}$$

where $e$ is the electronic charge and $T$ is the temperature.

Supplementary Figures S27 and S31 show the Seebeck coefficient $S$ evaluated at room temperature for different values of $E_F$, relative to the DFT-predicted Fermi energy $E_F^{DFT}$.

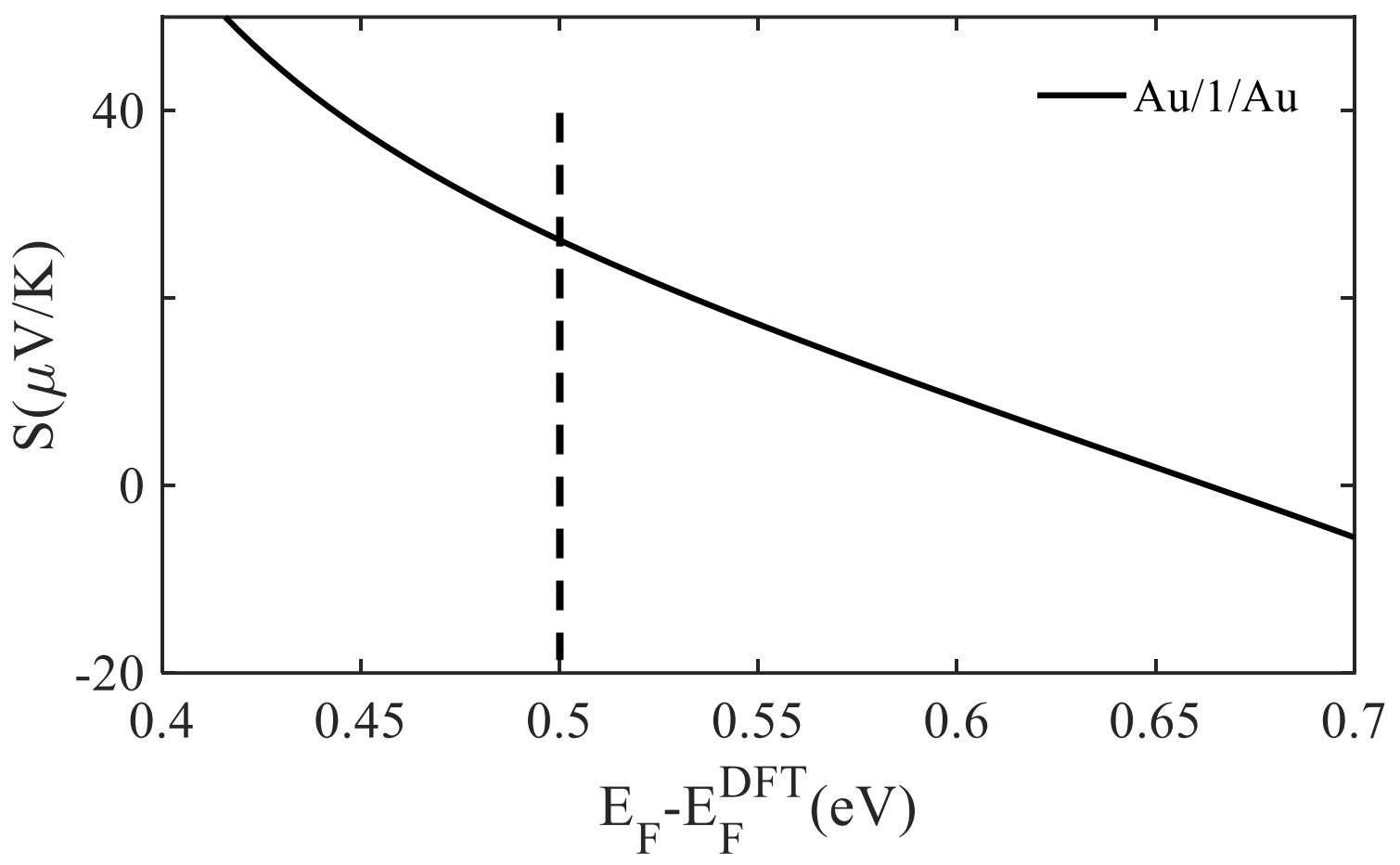

FIG S27. Seebeck coefficient S as a function of Fermi energy $E_F$ for **1,** (1,5-Di(4-(ethynylphenyl)thioacetate) anthracene.

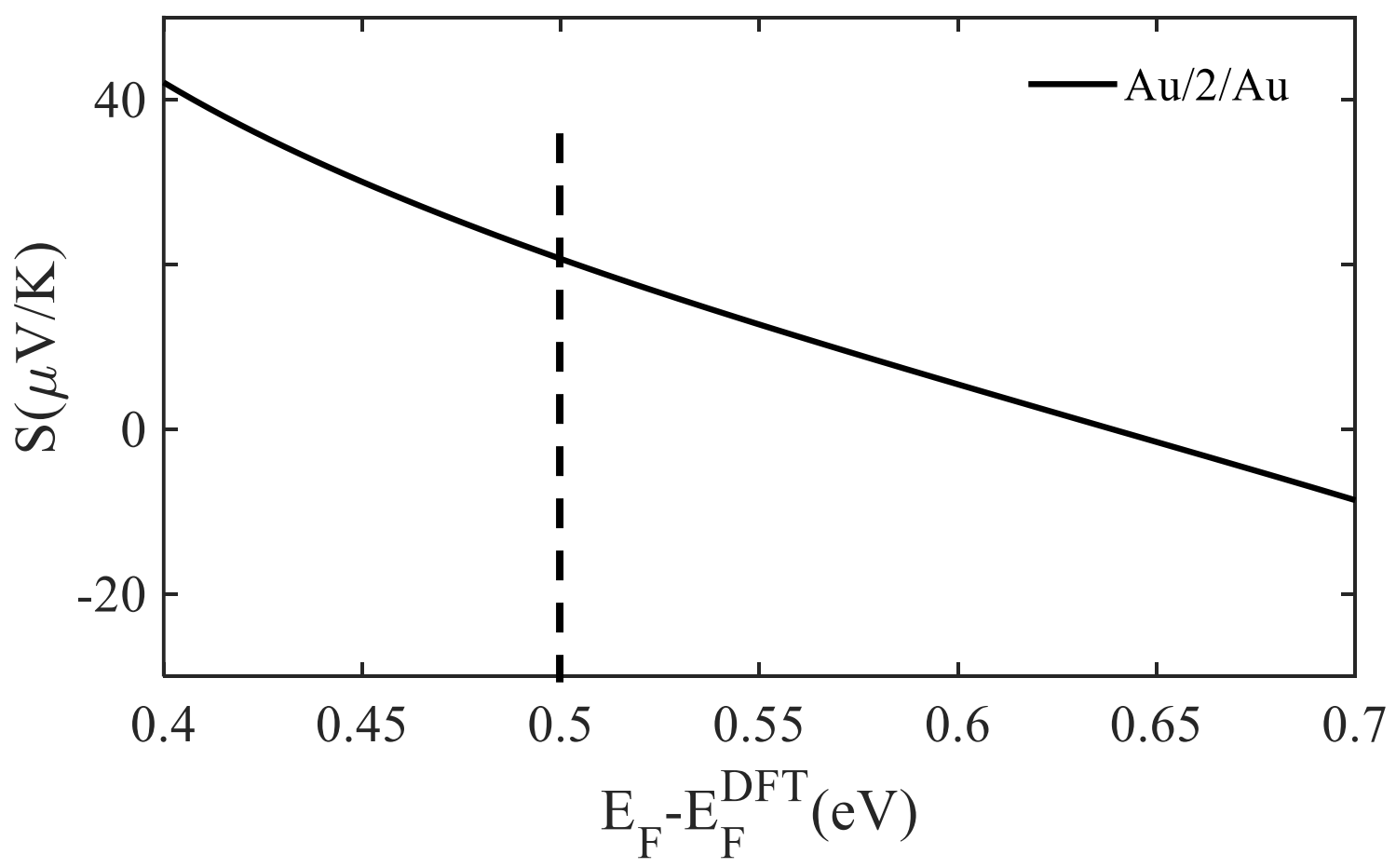

FIG S28. Seebeck coefficient S as a function of Fermi energy $E_F$ for **2**, (9,10-Di(4-(ethynylphenyl)thioacetate) anthracene.

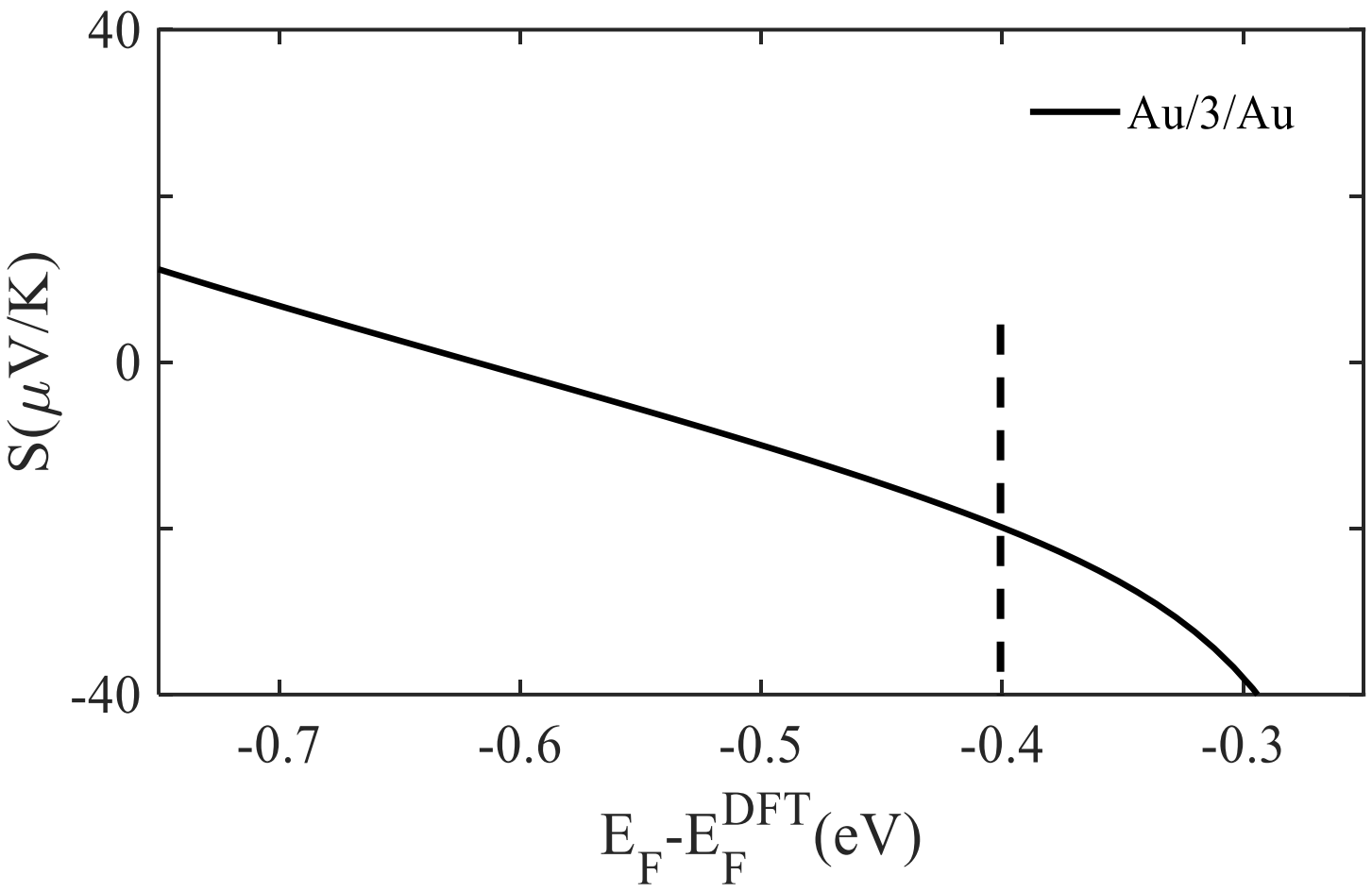

FIG S29. Seebeck coefficient S as a function of Fermi energy $E_F$ for **3**, (9,10-Di(4-ethynylthioanisole) anthracene.

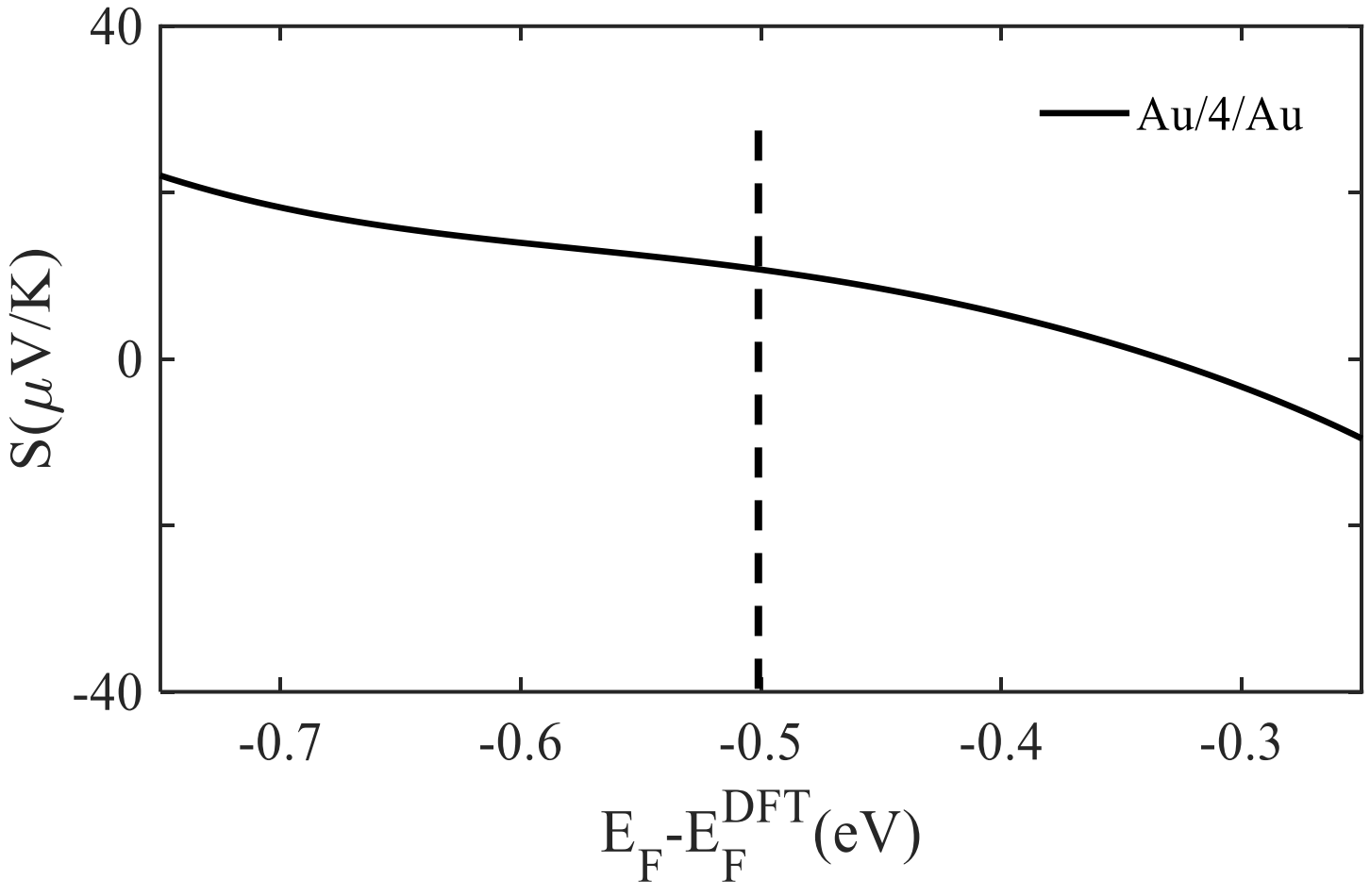

FIG S30. Seebeck coefficient S as a function of Fermi energy $E_F$ for **4**, (9-(4-(ethynylphenyl)thioacetate)-10-(4-ethynylpyridine)anthracene**.**

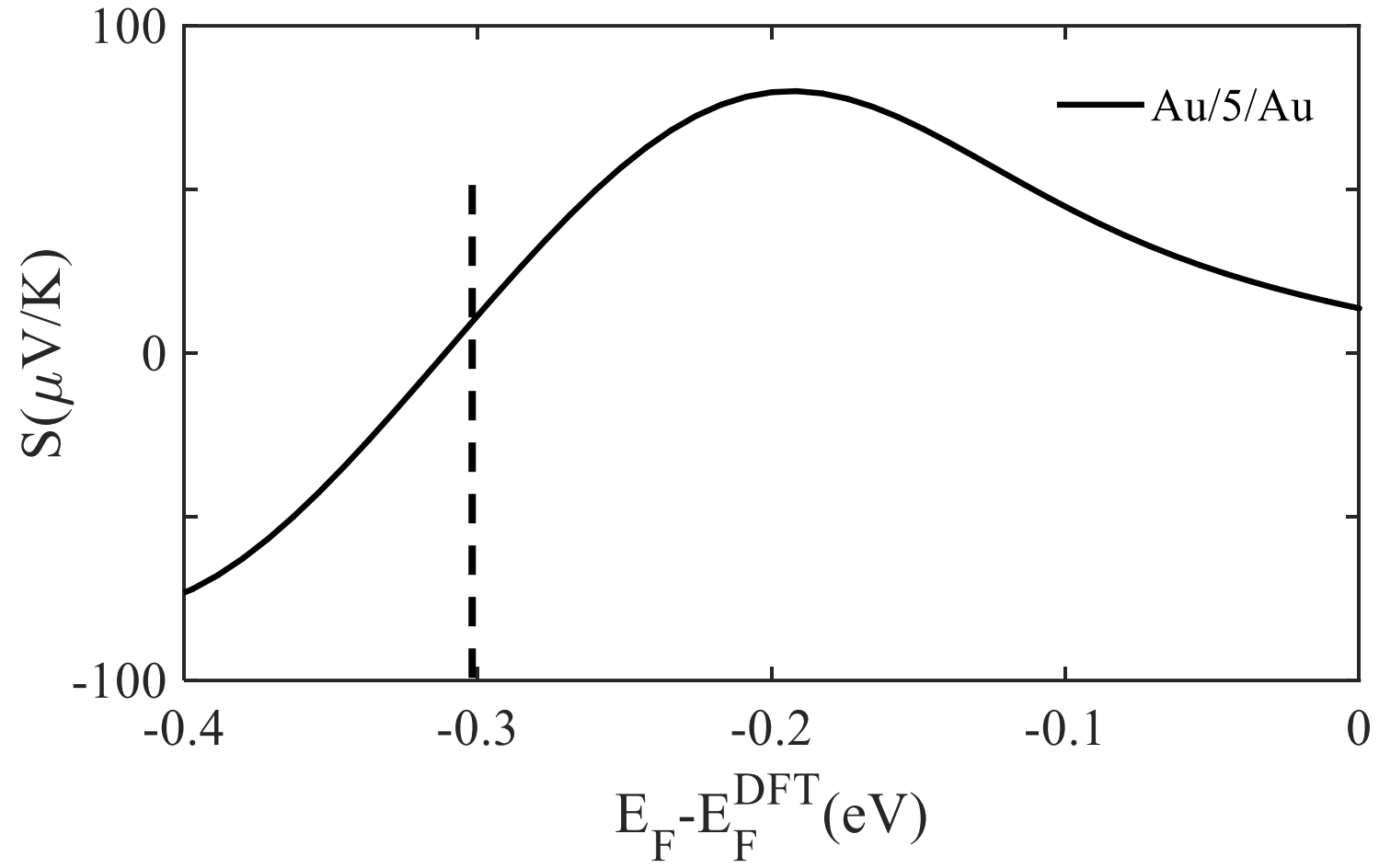

FIG S31. Seebeck coefficient S as a function of Fermi energy $E_F$ for **5**, (9-(4-ethynlthioanisole)-10-(4-ethynylpyridine) anthracene**.**

3.7 Comparison between transport properties obtained using Au-Au and Pt-Au electrodes.

Figs. S32, S34 and S36 (molecules 1, 2 and 4), show that the both junctions yield approximately the same HOMO-LUMO transmission curves (red and black). However, red curves (Pt-Au junction) are downshifted towards a lower energy by about 0.2 eV, reflecting their different electron affinities of Au=223 and Pt=205 kJ/mole. The 0.2 eV shift causes the slope of the two curves (red and black) of molecule 4, to switch from positive (black) to negative (red), as shown in the yellow-dotted rectangle box. Since the slopE of the two curves change then the Seebeck coefficient must switch sign and that is clearly shown in Fig. S37.

The case for molecules **1** and **2** is slightly different, where the 0.2 eV shift causes the slopes of the two curves (red and black), to remain unchanged or to become flat (zero, red), as shown in the yellow-dotted rectangle box. Since the slopes of the two curves are unchanged their the Seebeck coefficients remain positive or approach zero, as shown in Figs. S33 and S35 (molecules **1** and **2**). This behaviour reflects the nature of the anchor group and whether the molecule is symmetric or asymmetric.

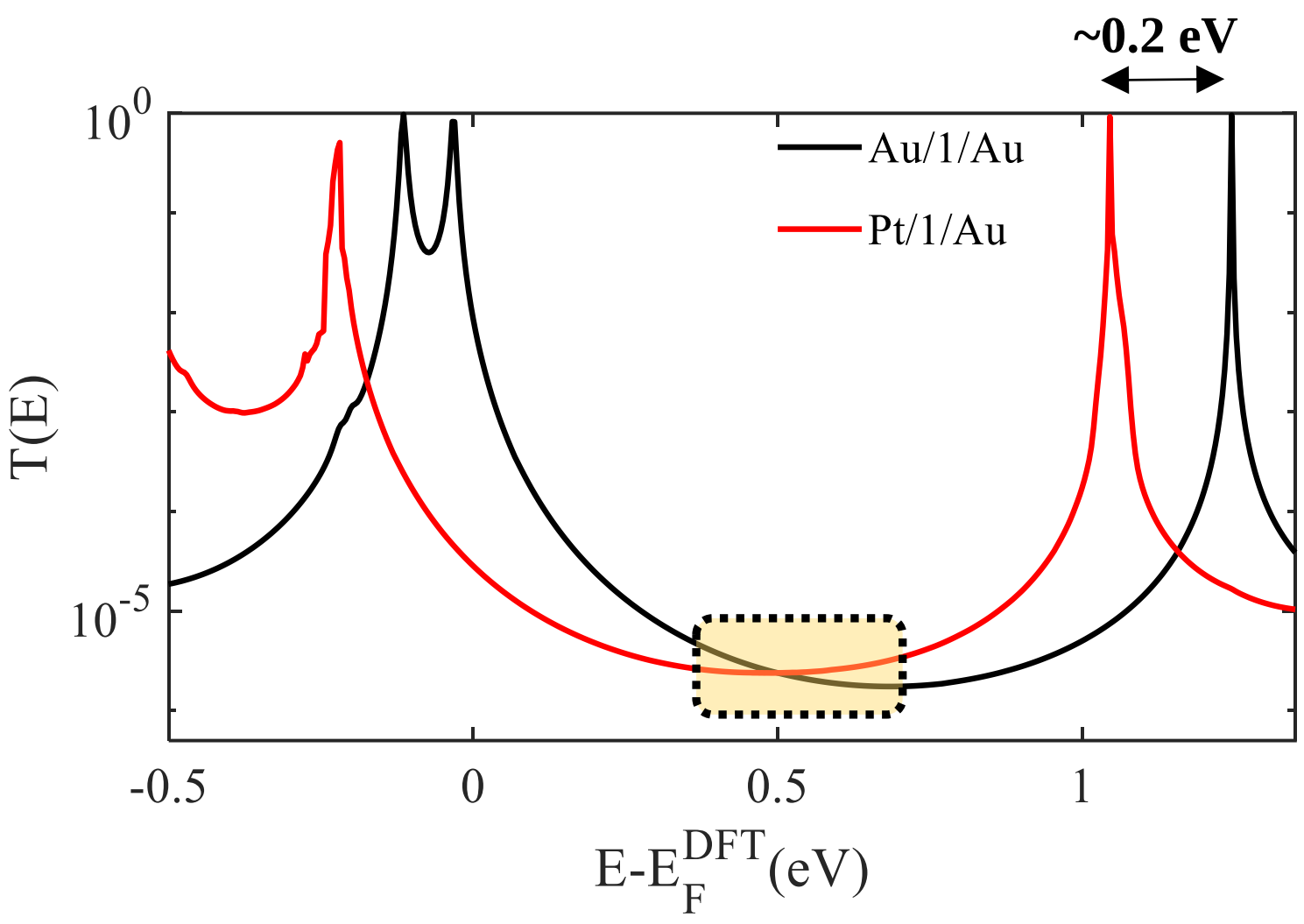

FIG S32. Zero bias transmission coefficients of **1** against electron energy *E* in two metallic junctions Au-Au (black lines), and Pt-Au (red lines).

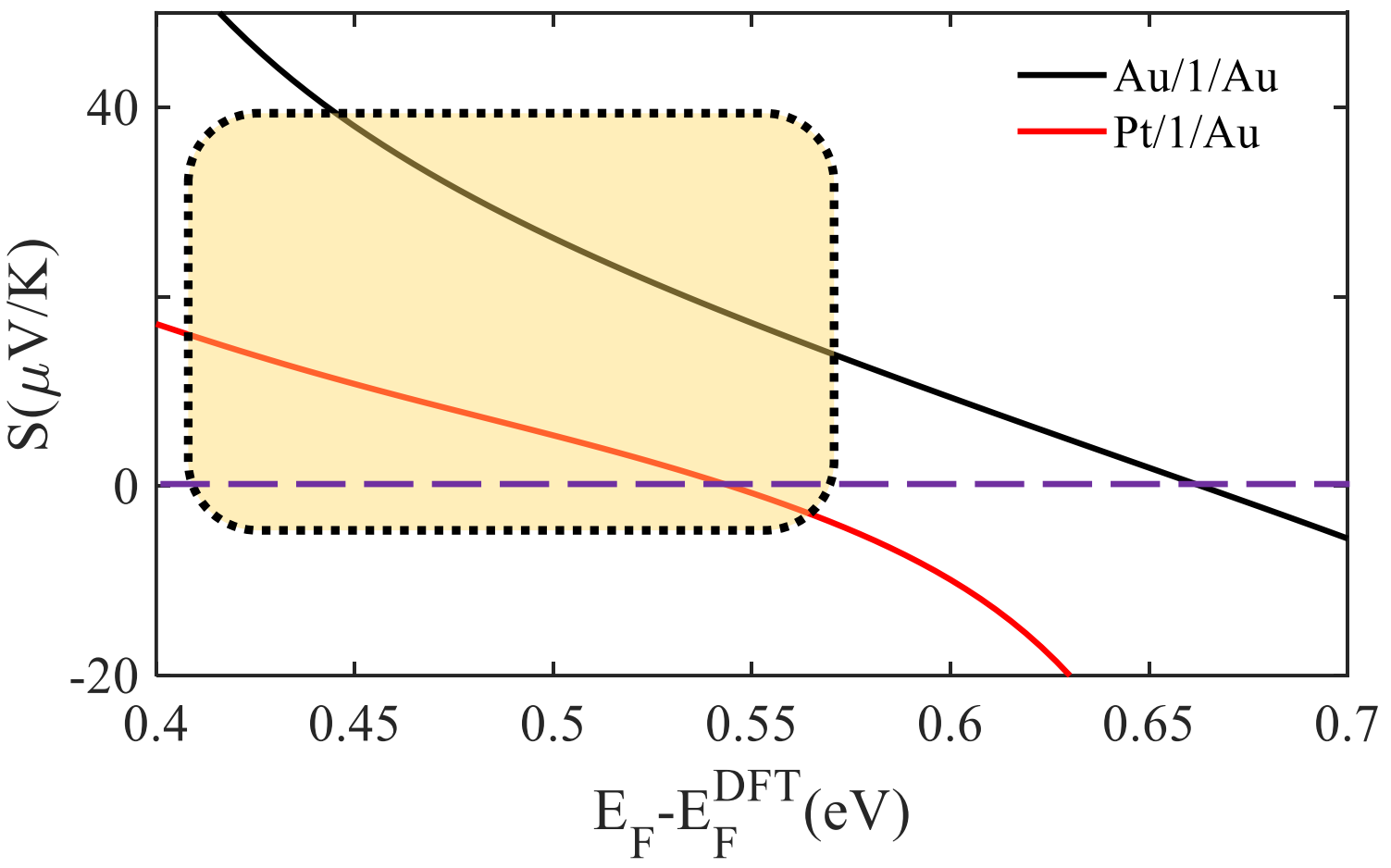

FIG S33. Seebeck coefficients S as a function of Fermi energy $E_F$ for anthracene **1** in two junctions Pt/**1**/Au and Au/**1**/Au. Mainly positive S of Au-Au junction (black curve), mainly positive S of Pt-Au junction (red curve).

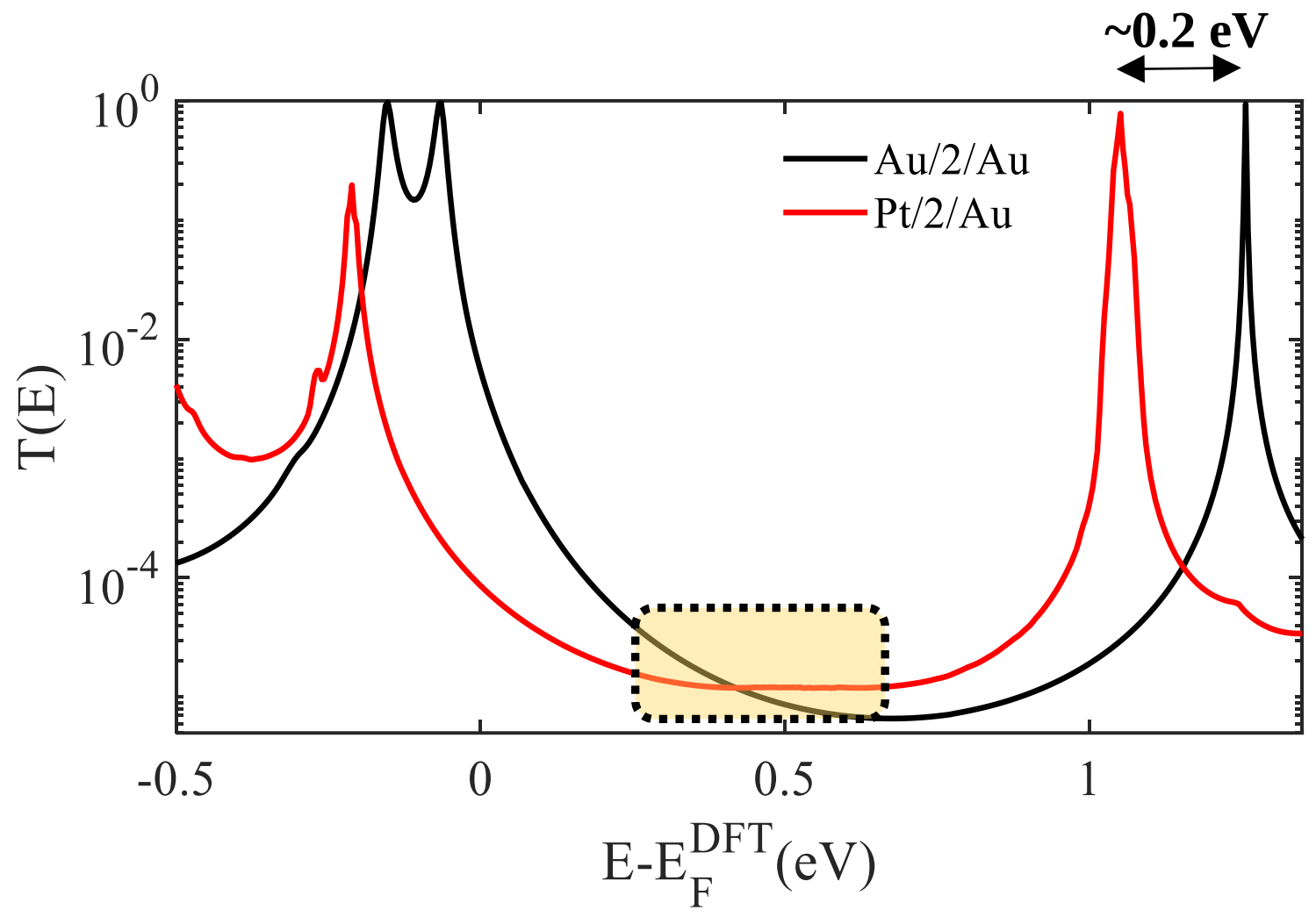

FIG S34. Zero bias transmission coefficients of **2** in two different junctions against electron energy E in two metallic junctions Au-Au (black lines), and Pt-Au (red lines).

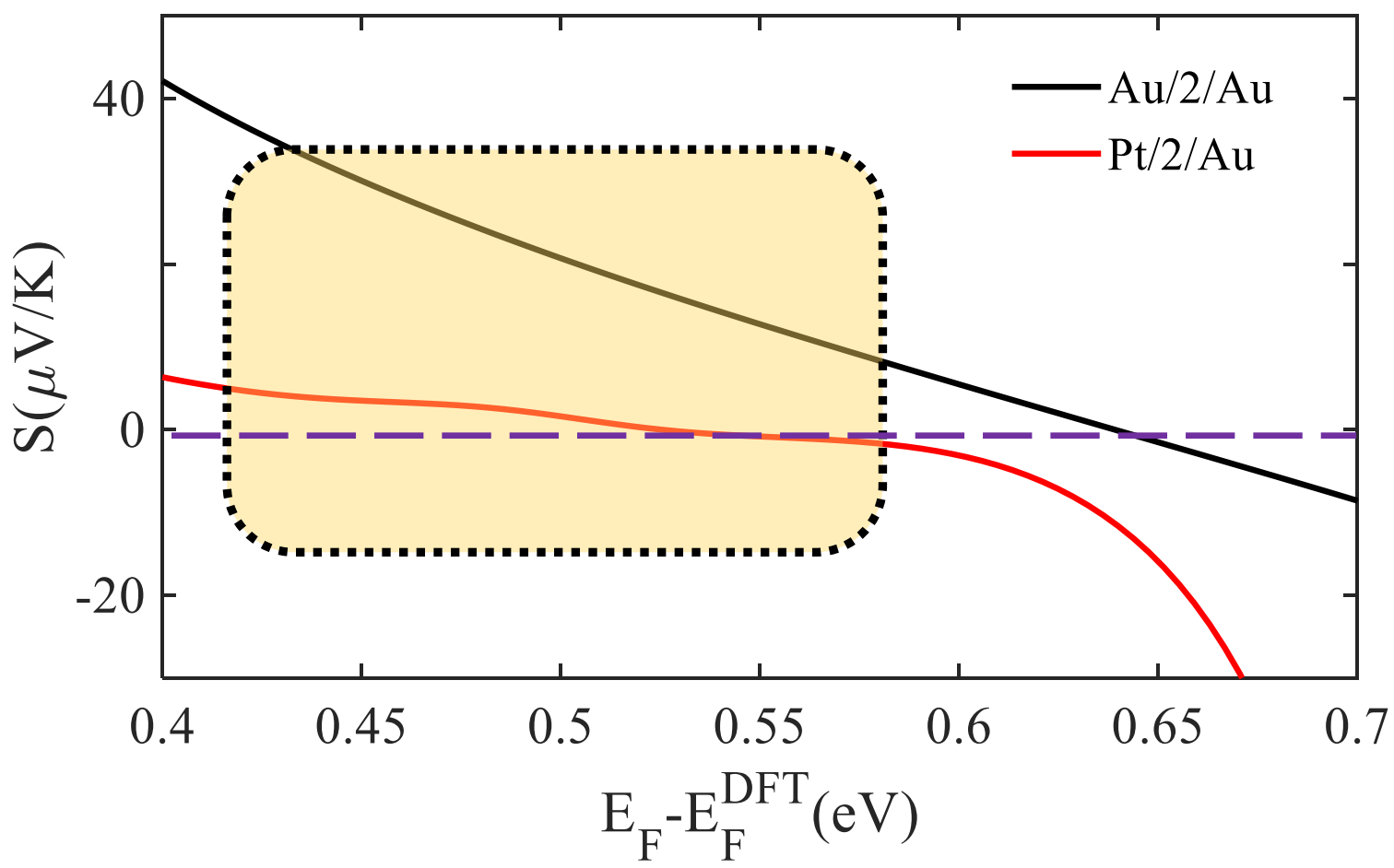

FIG S35. Seebeck coefficients S as a function of Fermi energy $E_F$ for anthracene **2** in two junctions Pt/**2**/Au and Au/**2**/Au. Mainly positive S of Au-Au junction (black curve), mainly positive S of Pt-Au junction (red curve).

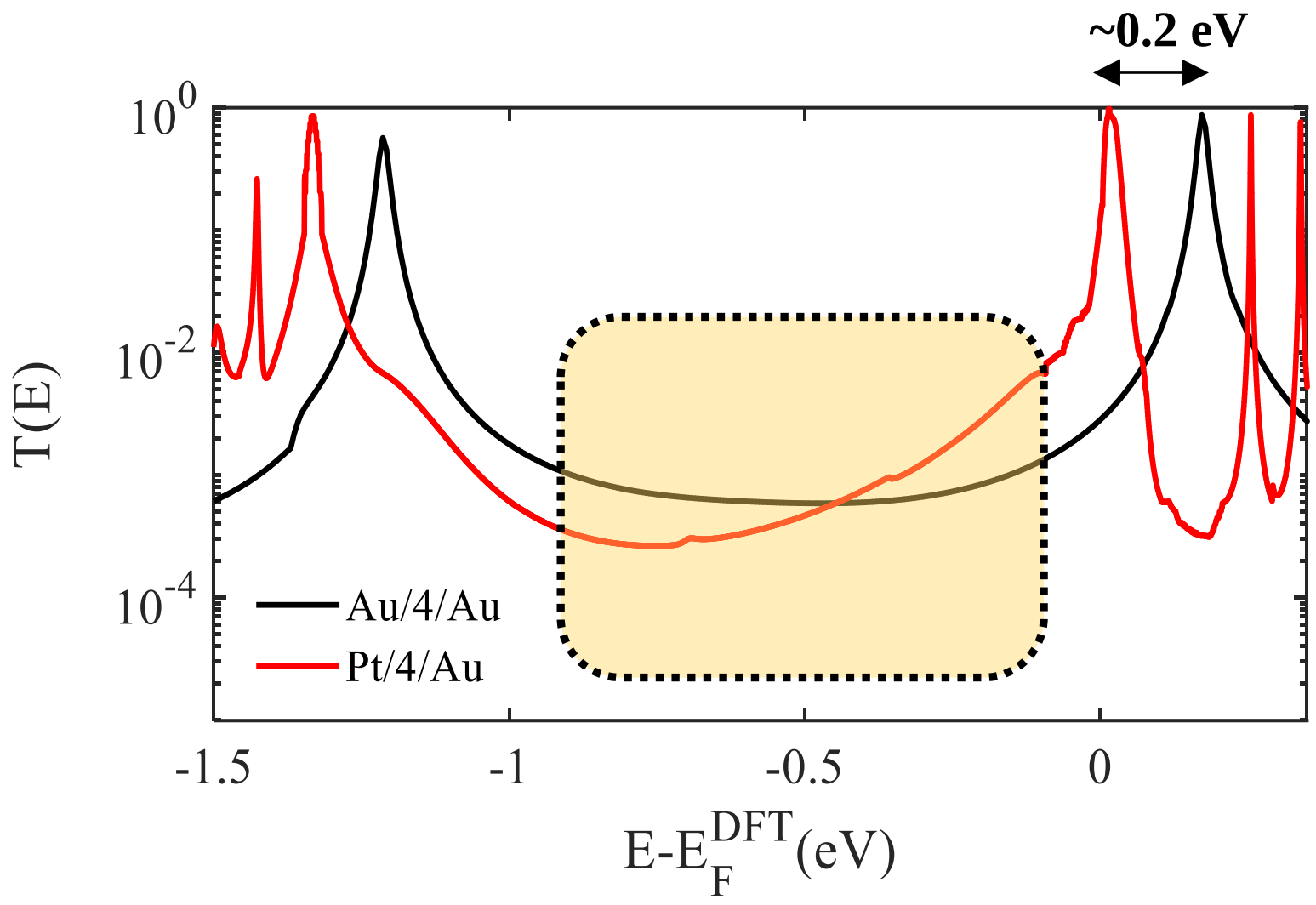

FIG S36. Zero bias transmission coefficients of **4** against electron energy *E* in two metallic junctions Au-Au (black lines), and Pt-Au (red lines).

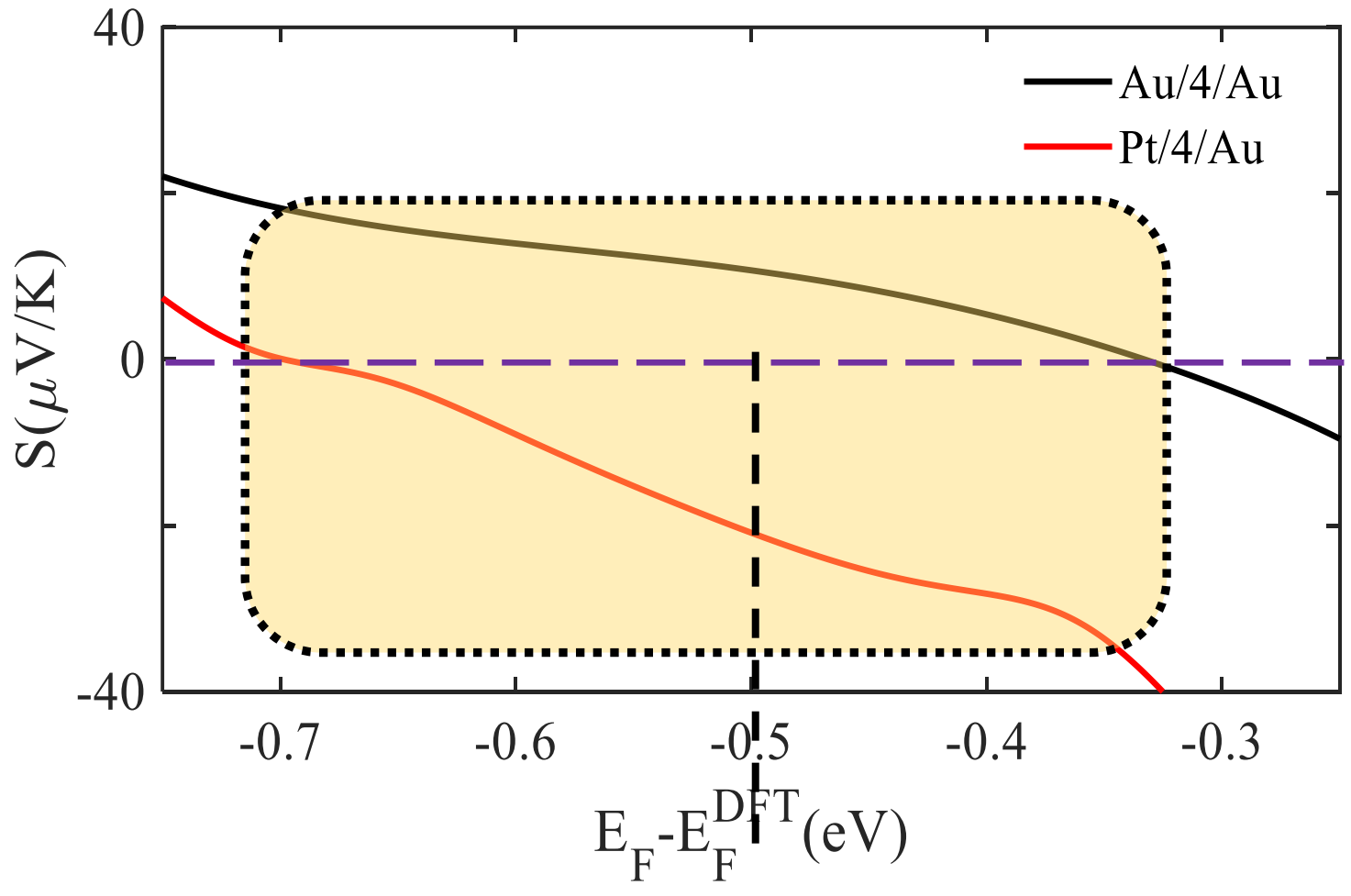

FIG S37. Seebeck coefficients S as a function of Fermi energy $E_F$ for anthracene **4** in two junctions Pt/**4**/Au and Au/**4**/Au. Mainly positive S of Au-Au junction (black curve), mainly negative S of Pt-Au junction (red curve).